\documentclass[a4paper,10pt]{article}
\pdfoutput=1 

\usepackage{jheppub} 

\usepackage[T1]{fontenc} 

\usepackage{amsfonts}
\usepackage{amsmath}
\usepackage{amsthm}
\usepackage{amssymb}
\usepackage{array}
\usepackage{dcolumn}
\usepackage{placeins}
\usepackage[svgnames]{xcolor}

\usepackage{graphics}
\usepackage{tikz}
\usepackage{graphicx}
\usetikzlibrary{positioning}
\usepackage{caption}
\usepackage{subcaption}
\usepackage{adjustbox}
\usepackage{gensymb}
\usepackage{breqn}
\usepackage{braket}
\usepackage{enumitem}
\usepackage{makecell}
\usepackage{ragged2e}
\usepackage{comment}
\usepackage{url}
\usepackage{placeins}
\usepackage{blindtext}
\usepackage{float}
\usepackage{hyperref}

\usepackage{fontawesome}

\preprint{ KIAS-P23062}

\newcommand\be{\begin{equation}}
\newcommand\ee{\end{equation}}
\newcommand\bea{\begin{eqnarray}}
\newcommand\eea{\end{eqnarray}}

\title{\boldmath Explaining ATOMKI, $(g-2)_\mu$, and MiniBooNE anomalies with light mediators in $U(1)_H$ extended model}

\author[a]{Sumit Ghosh}
\author[a]{and Pyungwon Ko}

\affiliation[a]{School of Physics, Korea Institute for Advanced Study, Seoul 02455, Korea}

\emailAdd{ghosh@kias.re.kr}
\emailAdd{pko@kias.re.kr}

\abstract{We consider $U(1)_H$ extensions of Type-I 2HDM plus a singlet scalar $\phi_H$, introducing a new Higgs doublet $H_2$ and a singlet $\phi_H$ charged under $U(1)_H$. The SM Higgs doublet $H_1$ as well as all the SM fermions and three right-handed singlet neutrinos, introduced to generate nonzero neutrino masses and mixings, are neutral under $U(1)_H$. We also introduce a SM singlet Dirac fermion, charged under $U(1)_H$ and utilize it as sterile neutrinos relevant to the MiniBooNE experiment. The $U(1)_H$ symmetry breaks due to the vacuum expectation values of $H_2$ and $\phi_H$, leading to the emergence of a light vector boson with a mass of approximately 17 MeV. This vector boson interacts with fermions through mass mixing and kinetic mixing process involving other neutral gauge bosons. Furthermore, alongside the light vector boson, another light scalar particle with a mass around 10--100 MeV may arise from scalar sector mixing. By utilizing the gauge couplings of the light vector boson and the Yukawa couplings of the light scalar, this model can  simultaneously provide an explanation for the Beryllium anomaly observed in the ATOMKI experiment, the anomalous magnetic moment of charged leptons and  the excess of electron-like events detected at the MiniBooNE experiment. }

\keywords{Light mediators, ATOMKI, MiniBooNE, $(g-2)_\mu$}

\begin{document} 
\maketitle
\flushbottom

\section{Introduction}

The landscape of fundamental particle physics currently stands at an interesting crossroad. On one hand, it is absolutely certain that the realm of physics must extend beyond the confines of the Standard Model (SM) despite its phenomenal success. Yet, on the other hand, the pursuit of these novel frontiers has proven to be very challenging. This quest for new physics beyond the SM finds a well-established foothold within the neutrino sector. The compelling oscillation data of neutrinos~\cite{Fukuda:1998mi,Ahmad:2002jz} unequivocally demand the inclusion of non-zero masses for at least two neutrinos. On the other hand, convincing evidence from cosmology reveals that a staggering $25\%$ of the Universe's matter exists in the form of elusive dark matter (DM)~\cite{Aghanim:2018eyx}, which is not described within the SM. While the vast majority of terrestrial experiments align with remarkable precision to the predictions of the SM, a select few results emerge as outliers. These interesting anomalies demand renewed understanding, challenging the comprehensive dominion of the SM.

The pursuit of physics beyond the SM (BSM) involves exploring a wide spectrum of particle masses and interactions. To reconcile BSM theories with the absence of current experimental findings, two avenues emerge: new particles might be extremely massive or possess very weak interactions with the SM. High-energy colliders like the Large Hadron Collider (LHC) play a pivotal role in directly detecting massive particles at the TeV scale. Complementary low-energy and/or high-intensity experiments focus on uncovering particles with lower mass and weak couplings. The prospect of new physics revealing itself through light and weakly  coupled  new states is gaining momentum, driven partly by the LHC's lack of significant signals at the TeV scale. In this context, low-energy and high-intensity experiments emerge as vital tools for investigation. Notably, a handful of intriguing hints pointing to deviations from the SM have surfaced in some of these low-energy experiments.

Rare nuclear transitions offer a promising avenue to detect new physics at lower energy scales, as they can be substantially influenced by BSM physics even if the coupling is extremely weak. While numerous nuclear transitions can be explored, those of particular interest involve the resonant production of excited nuclear states, followed by their decay through internal pair creation (IPC). In IPC, the excited nucleus emits a virtual photon that subsequently decays into an $e^+e^-$ pair. This high-statistics phenomenon serves as an excellent platform for investigating new physics with weak interactions. In recent times, the ATOMKI pair spectrometer experiment~\cite{Gulyas:2015mia} has reported an intriguing observation: an excess of 6.8 $\sigma$ in IPC decays of excited $^8$Be(18.15) nuclei~\cite{Krasznahorkay:2015iga, Krasznahorkay:2018snd}. This anomaly manifests as noticeable bumps in both the invariant mass and the angular opening of the $e^+e^-$ pairs. This stands in stark contrast to the predictions of electromagnetic (EM) IPC from virtual photons, which anticipate a smooth and rapidly declining distribution of opening angles. The presence of this anomalous bump aligns with expectations if a slowly moving vector boson with a mass of 17 MeV was produced on-shell via the emission of the excited nucleus, followed by its decay into $e^+e^-$ pairs. More recently, analogous anomalous outcomes have surfaced for IPC decays of excited $^4$He~\cite{Krasznahorkay:2019lyl, Krasznahorkay:2021joi} and $^{12}$C~\cite{Krasznahorkay:2022pxs} nuclei.

The magnetic dipole moment of the muon can be calculated and measured with exceptional precision, serving as a classic test of the SM~\cite{Keshavarzi:2021eqa}. Recently, the E989 experiment~\cite{Muong-2:2015xgu, Fienberg:2019ddu} at Fermi National Laboratory (FNAL) achieved the most accurate measurement of the muon's anomalous magnetic moment by combining data from its initial three runs~\cite{Muong-2:2023cdq}. This outcome agrees perfectly with the earlier measurement from Brookhaven National Laboratory (BNL)~\cite{Muong-2:2006rrc} and amplifies the world average discrepancy from the SM prediction to a significance of $5.1\sigma$. Ongoing efforts aim to clarify the existing theoretical landscape~\cite{Colangelo:2022jxc}, with an updated prediction incorporating comprehensive data from lattice calculations potentially leading to a narrower and less significant deviation. On the experimental front, an upcoming J-PARC experiment~\cite{Saito:2012zz} aims to cross-validate these findings. Potential explanations for this tension may emerge from quantum loop corrections linked to new electroweak (EW) scale particles around $100$ GeV, characterized by $\mathcal{O}(1)$ couplings, or from significantly lighter particles in the range of $10-100$ MeV, exhibiting $\mathcal{O}(10^{-4})$ couplings~\cite{Jaeckel:2020dxj}.

Neutrino oscillation experiments detect charged leptons resulting from (anti)neutrino charged current interactions with a target. These experiments utilize appearance measurements, seeking charged leptons with different flavors from the initially produced neutrinos. Among these experiments, MiniBooNE at FNAL~\cite{MiniBooNE:2008paa} was designed to detect electron (anti)neutrinos in a muon (anti)neutrino beam. It focused on identifying electron-like charged current quasi-elastic (CCQE) events originating from both neutrino and antineutrino modes. Interestingly, a significant surplus of events was detected in both modes. Recently, after 17 years of data collection, MiniBooNE updated their results~\cite{Aguilar-Arevalo:2020nvw}, revealing that the consistently observed excess in both neutrino and antineutrino modes now stands at a 4.8 $\sigma$ significance level. This establishes it as one of the most statistically significant anomaly in the neutrino sector. The traditional explanations, which leaned on oscillations driven by mixing with a new eV-scale neutrino, are now greatly challenged and possibly ruled out. This prompts the consideration of new physics in the neutrino sector as a potential solution to this excess.

The aforementioned anomalies could find an explanation through a new physics scenario that links to the SM via lightly connected mediator particles. The idea of such weakly coupled particles has been a plausible theoretical concept for a while. A compelling possibility for BSM new physics is the existence of new gauge groups. One scenario involves the potential survival of a remnant of grand unification at lower energies, such as the $U(1)_{T3R}$ gauge group~\cite{Dutta:2019fxn, Dutta:2022qvn}. Other prospects encompass the gauging of specific combinations of the accidental $B$ and $L_i$ global symmetries of the SM~\cite{Foot:1990mn, He:1990pn, He:1991qd}. If these symmetries break spontaneously at lower energies, they also exhibit weak coupling. Another intriguing and well-founded option is $U(1)_H$, initially proposed to tackle the significant flavor-changing neutral current (FCNC) issue in the two Higgs doublet model (2HDM)~\cite{Ko:2012hd}. In general 2HDMs, sizeable FCNC could occur through neutral Higgs bosons, often regarded as a phenomenological challenge. Glashow and Weinberg introduced criteria to address this problem, typically involving a softly broken $Z_2$ symmetry, yielding Type-I, II, X, Y 2HDMs~\cite{Glashow:1976nt, Branco:2011iw}. Some years ago, one of the authors of this paper proposed  implementing the $Z_2$ symmetry into a continuous $U(1)_H$ gauge symmetry, either eliminating or replacing the usual soft $Z_2$ breaking through spontaneous $U(1)_H$ breaking~\cite{Ko:2012hd}. In the former scenario, the massless Goldstone boson is absorbed by the $U(1)_H$ gauge boson, resulting in the absence of the typical pseudoscalar in the scalar spectra. In the latter case, introducing an SM singlet scalar $\phi_H$ with a nonzero $U(1)_H$ charge allows $U(1)_H$ symmetry breaking through the nonzero vacuum expectation value (vev) of $\phi_H$, inducing the expected soft $Z_2$ breaking effects. These $U(1)_H$ extensions of 2HDMs have been explored in various high-energy contexts, such as Higgs phenomenology~\cite{Ko:2013zsa}, dark matter within the inert 2HDM~\cite{Ko:2014uka}, and $E_6$-motivated leptophobic $Z_H$~\cite{Ko:2015fxa,Ko:2016lai}. In this paper, we explore the low-energy manifestation of $U(1)_H$.

The intriguing question arises: can any simple extension of the SM explain all these observations concurrently?  There have been attempts to address one or two of these anomalies at a time within a model framework in the literature~\cite{DelleRose:2017xil, Bertuzzo:2018itn, Datta:2020auq, Dutta:2020scq, Ballett:2019pyw}, whereas we aim  to explain all three simultaneously.  To achieve this, we propose a straightforward expansion of the SM based on the $U(1)_H$ gauge group. This extension introduces an additional SM scalar doublet and a singlet, both charged within the $U(1)_H$ framework. The expansion of the Higgs sector remains uncomplicated yet well-motivated, tied to the electroweak sector of the SM. The combined vev of the two doublets result in the spontaneous breaking of electroweak symmetry, while the vev of the SM singlet scalar and $U(1)_H$ charged doublet lead to the symmetry breaking of $U(1)_H$. Moreover, three right-handed neutrinos are introduced, with the further assumption that all fermions remain neutral under the $U(1)_H$ gauge group. Interactions between the new gauge boson and fermions arise from the interplay of mixing among the neutral gauge bosons. The most notable feature of the model is the emergence of a light scalar particle with a mass around 10-100 MeV, stemming from scalar sector mixing, and a light neutral vector boson with a mass around 17 MeV, resulting from neutral gauge boson sector mixing. Additionally, the presence of sterile neutrinos enables the realization of small neutrino masses via the type I seesaw mechanism within this framework. The lightest sterile neutrino could also serve as a viable dark matter candidate. The parameter space governing the light scalar and vector boson is outlined by various low-energy experiments. Within this space, we identify an allowable range to employ the light vector boson in clarifying the Beryllium anomaly. Furthermore, the excess events observed in the MiniBooNE experiment can primarily be attributed to the light scalar particle and a heavy sterile neutrino with a mass around 400 MeV. While both the light vector and scalar contribute to the muon's magnetic moments with opposing signs, fine-tuning between them allows for the attainment of accurate values.

The subsequent sections of this paper are structured as follows. In Section~\ref{sec:model}, a comprehensive overview of the model is provided. Specifically, Section~\ref{sec:scalars} delves into the scalar sector of the model, while Section~\ref{sec:fermions} elaborates on the Yukawa interactions between fermions and scalars. The  neutrino masses and mixing are explored in Section~\ref{sec:neutrinos}, followed by an in-depth discussion of gauge boson mass, mixings, and fermion-gauge interactions in Section~\ref{sec:gauge}. Turning our attention to specific phenomena, Section~\ref{sec:ATOMKI} undertakes an examination of the parameter space associated with the anomalous result observed in the ATOMKI experiment. An exploration of anomalous magnetic moments is undertaken in Section~\ref{sec:g2}, while Section~\ref{sec:MB} delves into the excess observed in the MiniBooNE experiment. To consolidate our findings, a concise summary is presented in Section~\ref{sec:summary}.

\section{Model Description} \label{sec:model}

We extend the SM gauge symmetry with a new abelian gauge group  $U(1)_H$ and additional new particles. A summary of the charges of the new particles in the model is given in Table.~\ref{tab:charges}.  Note that the new gauge symmetry does not contribute to electric charge. This new gauge symmetry is supposed to be broken at a relatively low energy scale by the condensations of two new scalar fields, $\phi_H$, which is charged under $U(1)_H$ but singlet under the SM gauge symmetry, and new Higgs doublet $H_2$ charged under $U(1)_H$.  In the following, we will briefly describe the masses and interaction terms for the scalars, fermions, and gauge bosons.

\begin{table}[tbp]
\centering
\begin{tabular}{c|llllll}
\hline
Symmetry Group & $H_1$ &$H_2$ &$\phi_H$&$n^\prime_R$&$\psi^\prime_L$&$\psi^{\prime c}_R$ \\

&& &&&&\\

\hline 
$SU(2)_L$ & 2 &2 &1&1&1&1 \\

&& &&&&\\

$U(1)_Y$ & 1/2&1/2 &0&0&0&0 \\

&& &&&&\\

$U(1)_H$ & 0&1 &$-1$&0&1& $-1$ \\

\hline
\end{tabular}
\captionsetup{justification   = RaggedRight,
             labelfont = bf}
\caption{\label{tab:charges} The quantum numbers of the particles under the gauge group $SU(2)_L \times U(1)_Y \times U(1)_H$ are reported. The other SM fermions have the usual charges, and none of them have charges under the new gauge group $U(1)_H$.  }
\end{table}

\subsection{Scalar sector} \label{sec:scalars}
The scalar sector comprises two SM doublet fields, denoted as $H_1$ and $H_2$, as well as one SM singlet field, $\phi_H$. These scalar fields possess following quantum numbers within the context of the $SU(2)_L \times U(1)_Y \times U(1)_H$ gauge symmetry:\begin{equation}
H_1\sim (2,1/2,0),~~~~~H_2\sim (2,1/2,1),~~~~~\phi_H\sim (1,0,-1)~,~\,
\end{equation} 
The most comprehensive, renormalizable, and CP-conserving scalar potential can be expressed in the following manner: 
\bea \label{eq:potential} 
V ( H_1 , H_2 , \phi_H ) &=& m^2_{1} H_1^\dagger H_1 + m^2_{2} H_2^\dagger H_2 + m_\phi^2\phi_H^\dagger \phi_H \nonumber \\ &~& +\frac{\lambda_1}{2} (H_1^\dagger H_1)^2  + \frac{\lambda_2}{2} (H_2^\dagger H_2)^2 + \frac{\lambda_\phi}{2} (\phi_H^\dagger \phi_H)^2  + \lambda_3 H_1^\dagger H_1H_2^\dagger H_2  + \lambda_4 H_1^\dagger H_2 H_2^\dagger H_1 \nonumber \\ &~&+  \lambda_{1 \phi} \phi_H^\dagger \phi_H H_1^\dagger H_1  + \lambda_{2 \phi} \phi_H^\dagger \phi_H H_2^\dagger H_2 \nonumber\\ &~& +m_{12 \phi} \left( H_1^\dagger H_2 \phi_H + \phi_H^\dagger H_2^\dagger H_1  \right) 
\eea 

In general, all three scalar fields can obtain nonzero vev's. Therefore, we take $\braket{H_1} =\left( 0, v_1/\sqrt{2} \right)$, $\braket{H_2} = \left( 0,v_2/\sqrt{2} \right)$, and $\braket{\phi_H} = v_\phi/\sqrt{2}$. The vevs $\braket{H_1} $ and $\braket{H_2}$ are responsible for electroweak symmetry breaking, where $v_1^2 + v_2^2 = v^2 = 246^2$~GeV$^2$, and $\tan \beta = v_2/v_1$. The vev of $\phi_H = v_\phi/\sqrt{2}$ breaks the new gauge symmetry of $U(1)_{H}$. The scale of $v_\phi$ can be chosen such that the new vector gauge boson has a very small mass, preferably in the MeV scale keeping 17 MeV $Z'$ for the ATOMKI anomaly in mind. After the symmetry breaking the scalar fields can be written as, \begin{eqnarray}
	 &H_1 &\sim   \left( \begin{array}{c}  {G}^+  \\  \frac{1}{\sqrt{2}}(v_1 +\rho_1+iG_0)  \end{array} \right),~~~~~~~~~~~~~~~~~~~~~~~~~~ H_2 \sim   \left( \begin{array}{c}  {\phi_2}^+ \nonumber \\  \frac{1}{\sqrt{2}}(v_2+\rho_2+i\eta_2)  \end{array} \right),~~~~\\&\phi_H&\sim \frac{1}{\sqrt{2}}(v_\phi+\rho_\phi+i G_{0 \phi})~.~\,
\end{eqnarray}

The extremization of the tree level scalar potential leads to the following conditions: 
\bea m_1^2 +\frac{\lambda_1 v_1^2}{2} +\frac{(\lambda_3 +\lambda_4) v_2^2}{2} +\frac{\lambda_{1\phi} v_\phi^2}{2} &=& -\frac{m_{12\phi} v_2 v_\phi}{\sqrt{2} v_1} \\ m_2^2 +\frac{\lambda_2 v_2^2}{2} +\frac{(\lambda_3 +\lambda_4) v_1^2}{2} +\frac{\lambda_{2\phi} v_\phi^2}{2} &=& -\frac{m_{12\phi} v_1 v_\phi}{\sqrt{2} v_2} \\ m_\phi^2 +\frac{\lambda_\phi v_\phi^2}{2}  +\frac{\lambda_{1\phi} v_1^2}{2}++\frac{\lambda_{2\phi} v_2^2}{2} &=& -\frac{m_{12\phi} v_1 v_2}{\sqrt{2} v_\phi}\eea

There are a total of 10 degrees of freedom (d.o.f.), out of which 4 ($G^\pm$, $G_0$, and $G_{0 \phi}$) will be absorbed by the massless gauge bosons to become massive. The remaining d.o.f. correspond to physical scalar particles in the model. Specifically, 2 d.o.f. are for the singly charged scalar $h^{\pm}$ (coming from $\phi_2^\pm$), 1 is for the pseudo-scalar, which we denote as $s$ (arising from $\eta_2$), and the remaining 3 d.o.f. correspond to the physical scalar particles $h, h_1, h_2$. These arise from the mixing of the three CP-even states, namely, $\rho_1, \rho_2$, and $\rho_\phi$.

The charged scalar mass square is given by,
\bea m^2_{h^{\pm}}&=&  -\frac{m_{12\phi} v_1 v_\phi}{\sqrt{2} v_2} - \frac{\lambda_4  v_1^2}{2} \eea where we have utilized the conditions arising from the extremization to simplify this. The pseudo-scalar mass square term is \bea m_s^2 &=& -\frac{m_{12\phi} v_1 v_\phi}{2\sqrt{2} v_2} \eea

Our primary focus lies in the mixing of the CP-even scalars, as this process leads to the emergence of the desired light scalar field. The three neutral CP-even components undergo mixing amongst themselves, described by the mass square matrix, 
\begin{equation}M^2_\rho = \left( \begin{array}{ccc}\lambda_1 v_1^2-\frac{m_{12\phi} v_2 v_\phi}{\sqrt{2} v_1} &(\lambda_3+\lambda_4) v_1 v_2 +\frac{m_{12\phi} v_\phi}{\sqrt{2}}&\lambda_{1\phi} v_1 v_\phi +\frac{m_{12\phi} v_2}{\sqrt{2}} \\ (\lambda_3+\lambda_4) v_1 v_2 +\frac{m_{12\phi} v_\phi}{\sqrt{2}}& \lambda_2 v_2^2-\frac{m_{12\phi} v_1 v_\phi}{\sqrt{2} v_2}&\lambda_{1\phi} v_2 v_\phi +\frac{m_{12\phi} v_1}{\sqrt{2}} \\ \lambda_{1\phi} v_1 v_\phi +\frac{m_{12\phi} v_2}{\sqrt{2}} &\lambda_{1\phi} v_2 v_\phi +\frac{m_{12\phi} v_1}{\sqrt{2}} &\lambda_\phi v_\phi^2-\frac{m_{12\phi} v_1 v_2}{\sqrt{2} v_\phi} \end{array} \right) ~.~\, \end{equation}

This mass square matrix can be diagonalized as $R^T M^2_\rho R$ using a $3 \times 3$ orthogonal matrix $R_{3\times 3}$ that can be parametrized using three rotation angles. 
\begin{eqnarray} \label{3x3mixing} R_{3\times 3} &=& \left( \begin{array}{ccc} r_{11} & r_{12} & r_{13} \\ r_{21} & r_{22} & r_{23} \\ r_{31} & r_{32} & r_{33} \end{array} \right) ~,~\,\end{eqnarray} where $r_{ij}$'s are the functions of $sine$ and $cosine$ of the three rotation angles.
The physical states can be defined as $R^T (\rho_1, \rho_2, \rho_\phi)$. The interaction states can be expanded as superposition of the physical states as, 
\begin{eqnarray} \rho_1 &=& r_{11} h + r_{12} h_2 + r_{13} h_1 ~,~\,\nonumber\\
\rho_2 &=&  r_{21} h + r_{22} h_2 + r_{23} h_1 ~,~\,\nonumber\\ 
\rho_\phi &=&  r_{31} h + r_{32} h_2 + r_{33} h_1~.~\,\end{eqnarray}

Here $h$ can be identified as the SM Higgs field; $h_2$ can be heavy such that it satisfy the LHC bounds. And we want a scenario where $h_1$ is a light scalar with mass ~$\mathcal{O}(10)$~MeV, which will play key roles to explain  the muon $(g-2)$ and the MiniBooNE excess (see Secs. 4 and 5, respectively for details).

We generate a phenomenologically interesting scalar mass spectrum by selecting appropriate values for the parameters in Eq.~(\ref{eq:potential}) (see for example Ref.~\cite{Dutta:2020scq, Dutta:2022fdt} for similar treatment). Specifically, we choose mass parameters of the order of $\sim \mathcal{O}(100)$~GeV and dimensionless coupling parameters around $\sim \mathcal{O}(0.1)$. As an example, we consider the following vacuum expectation values (vevs): $v_1 = 238$ GeV; $v_2=62$ GeV; and $v_\phi=10$ GeV. This parameter selection leads to the following scalar spectrum: $m_{h} = 125$ GeV; $m_{h_2}= 500$ GeV; and $m_{h_1} = \mathcal{O}(10-100)$ MeV. 

The SM Higgs boson $h$ can decay into a pair of $h_1$ particles, contributing to the invisible decay channel of the SM Higgs. Experimental searches for such invisible decay channels at the LHC have yielded null results so far, constraining the decay branching fraction to $Br(h \rightarrow invisible ) < 0.15$ at a $95\%$ confidence level~\cite{CMS:2023sdw, ATLAS:2023tkt}. For the chosen parameter set, the coupling $h h_1 h_1$ is of the order of $\mathcal{O}(0.1)$, and the corresponding branching fraction is approximately $\mathcal{O}(0.01)$.

\subsection{Fermion sector} \label{sec:fermions}

In the fermion sector we introduce three SM singlet right handed fields $n_R^\prime$ which are singlet under $U(1)_H$ as well. In addition to this we introduce a pair of Weyl fermion $\psi_{L,R}$ which are charged under $U(1)_H$ with charge 1  but singlet under SM. The new fermionic fields posses following quantum numbers within the context of the $SU(2)_L \times U(1)_Y \times U(1)_H$ symmetry:\begin{equation}
n^\prime_R\sim (1,0,0),~~~~~\psi^\prime_L\sim (1,0,1),~~~~~\psi^{\prime c}_R \sim (1,0,-1)~,~\,
\end{equation}  The charges of $\psi^\prime_L$ and $\psi^\prime_R$ are chosen in such a way to cancel the anomalies in the model. The Lagrangian responsible for fermion masses and mixing can be given by,
\bea \label{eq:LYukawaint} -\mathcal{L}_{Y} &=& \bar{q}^\prime_{L_i}(y^\prime_{d})_{ij} d^\prime_{R_j} H_1 +\bar{q}^\prime_{L_i}(y^\prime_{u})_{ij} u^\prime_{R_j} \tilde{H_1}\nonumber\\ &~&+\bar{l}^\prime_{L_i}(y^\prime_{e})_{ij} e^\prime_{R_j} H_1 +\bar{l}^\prime_{L_i}(y^\prime_{n})_{ij} n^\prime_{R_j} \tilde{H_1}+ \frac{1}{2}\bar{n}^{\prime c}_{R_i} {M}^\prime_{ij}n^\prime_{R_j} + m_\psi \bar{\psi}^\prime_R \psi^\prime_L \nonumber\\ &~& + \bar{n}^{\prime c}_{R_i} (y^\prime_{1\psi})_i \psi_R^\prime \phi^\dagger_H + \bar{\psi}^\prime_L (y_{2\psi}^\prime)_jn^\prime_{R_j} \phi_H  +  H.c.   \eea where primed fermions are interaction basis fermions; $i$ and $j$ are the family indices, $i,j= 1,2,3$. The first two terms give down-type and up-type quark masses and their interactions with physical scalars. The next four terms give charged lepton masses, Dirac mass term for neutrinos, Majorana mass term for right handed neutrinos, and mass term for the fermion $\psi$ respectively.  Note that there is no Yukawa type term for the second doublet $H_2$ as it is charged under the new gauge group. Therefore our model is an $U(1)_H$ extension of the Type I 2HDM. Though dimension five non-renormalizable term could be possible, but we do not consider them. In general, all the Yukawa matrices are $3 \times 3$ complex matrices, but for simplicity we assume that they are real.

The Yukawa matrices can be diagonalized using biunitary transformation and the Majorana mass matrix can be diagonalized using unitary transformation as follows, 
\begin{eqnarray} \label{def:diag}y_d &= &U^\dagger_{d_{L}} y^\prime_{d} U_{d_{R}} ,~~~~~\mbox{with}~~~(y_{d})_{ij} = (y_{1d})_{ii} \delta_{ij} ~,~\,\\
 y_u &=&U^\dagger_{u_{L}} y^\prime_{u} U_{u_{R}} ,~~~~~\mbox{with}~~~(y_{u})_{ij} = (y_{u})_{ii} \delta_{ij} ~,~\,\\
 y_e &=&U^\dagger_{e_{L}} y^\prime_{e} U_{e_{R}} ,~~~~~\mbox{with}~~~(y_{e})_{ij} = (y_{1e})_{ii} \delta_{ij} ~,~\,\\
 y_n &=&U^\dagger_{\nu_{L}} y^\prime_{n} U_{n_{R}} ,~~~~~\mbox{with}~~~(y_{n})_{ij} = (y_{1n})_{ii} \delta_{ij} ~,~\,\\
 M &=& U^\dagger_{n_{R}}M^\prime U_{n_{R}} ,~~~~~\mbox{with}~~~M_{ij}=M_{ii} \delta_{ij}~,~\,
\end{eqnarray} We further define, \bea y_{1\psi} &=& U^\dagger_{n_{R}} y^\prime_{1\psi} ~,~\, \\ y_{2\psi} &=&  y^\prime_{2\psi} U_{n_{R}}~,~\, \eea

Here $U$'s are $3\times 3$ Unitary matrices which can be used to define the fermions in the mass basis as follows, \bea \left(d_{{L/R}}\right)_i &=& \left(U^\dagger_{d_{L / R}}\right)_{ij} \left(d^\prime_{{L/R}}\right)_j ~,~\, \\ \left(u_{{L/R}}\right)_i &=& \left(U^\dagger_{u_{L / R}}\right)_{ij} \left(u^\prime_{{L/R}}\right)_j ~,~\, \\ \left(e_{{L/R}}\right)_i &=& \left(U^\dagger_{e_{L / R}}\right)_{ij} \left(e^\prime_{{L/R}}\right)_j ~,~\, \\ \left(\nu_{{L}}\right)_i &=& \left(U^\dagger_{\nu_{L}}\right)_{ij} \left(\nu^\prime_{{L}}\right)_j ~,~\,\\\left(n_{{R}}\right)_i &=& \left(U^\dagger_{n_{R}}\right)_{ij} \left(n^\prime_{{R}}\right)_j \\\eea

Now Eq.~(\ref{eq:LYukawaint}) can be rewritten as, 
\bea -\mathcal{L}_{Y} &=& {m_d}_i \bar{d}_i d_i+\frac{{m_d}_i}{v_1} \bar{d}_i d_i \rho_1  +{m_u}_i \bar{u}_i u_i+\frac{{m_u}_i}{v_1} \bar{u}_i u_i \rho_1  +{m_e}_i \bar{e}_i e_i+\frac{{m_e}_i}{v_1} \bar{e}_i e_i \rho_1 \nonumber\\ &~& + {m_{\nu_D}}_i (\bar{\nu}_{L_i} n_{R_i} + \bar{n}_{R_i} \nu_{L_i})+ \frac{{m_{\nu_D}}_i}{v_1}(\bar{\nu}_{L_i} n_{R_i} + \bar{n}_{R_i} \nu_{L_i}) \rho_1  \nonumber\\ &~&+ \frac{1}{2} M_i \left( \bar{n}^c_{R_i} n_{R_i} + \bar{n}_{R_i} n^c_{R_i}  \right) + m_\psi \bar{\psi} \psi  \nonumber\\ &~&   + \frac{(y_{1\psi})_i v_X}{\sqrt{2}} \left( \bar{n}^c_{R_i} \psi_R + \bar{\psi}_R n^c_{R_i}  \right)+ \frac{(y_{2\psi})_i v_X}{\sqrt{2}} \left( \bar{\psi}_{L} n_{R_i} + \bar{n}_{R_i} \psi_{L}  \right) \nonumber\\ &~& + \frac{(y_{1\psi})_i }{\sqrt{2}} \left( \bar{n}^c_{R_i} \psi_R + \bar{\psi}_R n^c_{R_i} \right) \rho_\phi+ \frac{(y_{2\psi})_i }{\sqrt{2}} \left( \bar{\psi}_{L} n_{R_i} + \bar{n}_{R_i} \psi_{L}  \right) \rho_\phi \eea where we have used the definitions: $m_{f_i} = (y_f)_{ii} \delta_{ij}v_1/\sqrt{2}$ with $f = d,u,e$; and $m_{{\nu_D}_i} = (y_n)_{ii} \delta_{ij}v_1/\sqrt{2}$. Now let us decompose the lagrangian for neutrino sector and rest of the fermions as follows, \be -\mathcal{L}_{Y} = -\mathcal{L}_{Y}(f=d,u,e) -\mathcal{L}_{Y}(\nu,n, \psi) \ee here the first term can be expressed compactly as, \be -\mathcal{L}_{Y}(f=d,u,e) = m_{f_i}\bar{f}_i f_i  + \bar{f}_i (y_{f\phi})_i f_i \phi \ee where $\phi = h,h_2,h_1$; and the couplings $(y_{f\phi})_i$ are defined as, \bea (y_{fh})_i &=& \frac{ m_{f_i}}{v_1} r_{11} \\ (y_{fh_2})_i &=& \frac{ m_{f_i}}{v_1} r_{12} \\ (y_{fh_1})_i &=& \frac{ m_{f_i}}{v_1} r_{13} \eea
Note that there is no tree level flavor changing scalar current and also there is no charged scalar current or pseudo scalar current. The scalars has couplings to SM fermions proportional to their mass.

The Lagrangian for neutrino sector can be written as \bea \label{eq:neutrinomass} -\mathcal{L}_{Y}(\nu,n,\psi) &=& {m_{\nu_D}}_i (\bar{\nu}_{L_i} n_{R_i} + \bar{n}_{R_i} \nu_{L_i})+\frac{1}{2} M_i \left( \bar{n}^c_{R_i} n_{R_i} + \bar{n}_{R_i} n^c_{R_i}  \right) + m_\psi \bar{\psi} \psi  \nonumber\\ &~&   + m_{{1n\psi}_i} \left( \bar{n}^c_{R_i} \psi_R + \bar{\psi}_R n^c_{R_i}  \right)+ m_{{2n\psi}_i} \left( \bar{\psi}_{L} n_{R_i} + \bar{n}_{R_i} \psi_{L}  \right)\nonumber\\ &~& + \frac{{m_{\nu_D}}_i}{v_1}(\bar{\nu}_{L_i} n_{R_i} + \bar{n}_{R_i} \nu_{L_i}) \rho_1  \nonumber\\ &~& + \frac{m_{{1n\psi}_i}}{v_X} \left( \bar{n}^c_{R_i} \psi_R + \bar{\psi}_R n^c_{R_i} \right) \rho_\phi+ \frac{m_{{2n\psi}_i}}{v_X} \left( \bar{\psi}_{L} n_{R_i} + \bar{n}_{R_i} \psi_{L}  \right) \rho_\phi \eea where we have defined, $m_{{1n\psi}_i} = (y_{1\psi})_i v_\phi/\sqrt{2}$ and $m_{{2n\psi}_i} = (y_{2\psi})_i v_\phi/\sqrt{2}$. The terms in the  first two  lines are responsible for neutrino masses and mixing while the term in the third and fourth lines give interaction of neutrinos with physical scalar particles.

\subsection{Neutrino masses and mixings} \label{sec:neutrinos}

Part of Eq.~(\ref{eq:neutrinomass}) that gives rise to neutrino masses and mixings is given by, \bea -\mathcal{L}^{mass}_{Y}(\nu,n) &=& {m_{\nu_D}}_i (\bar{\nu}_{L_i} n_{R_i} + \bar{n}_{R_i} \nu_{L_i}) + \frac{1}{2} M_i \left( \bar{n}^c_{R_i} n_{R_i} + \bar{n}_{R_i} n^c_{R_i}  \right)  + m_\psi \bar{\psi} \psi \nonumber\\ &~&+ m_{{1n\psi}_i} \left( \bar{n}^c_{R_i} \psi_R + \bar{\psi}_R n^c_{R_i} \right) + m_{{2n\psi}_i}\left( \bar{\psi}_{L} n_{R_i} + \bar{n}_{R_i} \psi_{L}  \right) \nonumber\\ &=& \frac{1}{2}\left(\bar{\nu}^c_{L_i} ~~~ \bar{n}_{R_i} ~~~ \bar{\psi}^c_L ~~~ \bar{\psi}_R  \right)~ M^\prime_\nu ~ \left( \begin{array}{c} \nu_{L_i} \\ n^c_{R_i} \\ \psi_L \\ \psi^c_R \end{array} \right) \nonumber\\ &=& \frac{1}{2} \left( \bar{\nu}_{a_i}~~~ \bar{\nu}_{s_j} \right) M_\nu \left( \begin{array}{c} \nu_{a_i} \\ \nu_{s_j} \end{array} \right) ~,~\, \eea where the  $8 \times 8$ neutrino mass matrix $M^\prime_\nu$ is given by, 
\be M^\prime_\nu = \left( \begin{array}{cccc} 0 & {{m^T_{\nu_D}}_i} & 0 &0 \\ {{m_{\nu_D}}_i} & M_i & m_{{2n\psi}_i} & m_{{1n\psi}_i}\\ 0 & m_{{2n\psi}_i} & 0&m_\psi\\ 0&m_{{1n\psi}_i} &m_\psi &0 \end{array} \right)   ~,~\, \ee
We diagonalize the neutrino mass matrix using an Unitary matrix $U_\nu$ as \be U^T_\nu M^\prime_\nu U_\nu = M_\nu = diag(m_1, m_2,....m_8) \ee We also define mass states as \be  \left( \begin{array}{c} \nu_{L_i} \\ n^c_{R_i} \\ \psi_L \\ \psi^c_R \end{array} \right) = U_\nu \left( \begin{array}{c} \nu_{a_i} \\ \nu_{s_j} \end{array} \right)\ee where $i=1,2,3$ and $j=1,2,3,4,5$. Therefore we have three active neutrino states which we identify as the SM light neutrinos while we have 5 sterile neutrinos which we identify as heavy neutrinos. 

The absolute masses of neutrinos remain unknown; instead, we have information about the differences in their squared masses, denoted as  $\Delta m^2_{ij} \equiv m_{\nu_{a_i}}^2-m_{\nu_{a_j}}^2$. In the case of the normal hierarchy, where $m_{\nu_{a_1}} \ll m_{\nu_{a_2}} < m_{\nu_{a3}}$, global fits of neutrino oscillation data~\cite{deSalas:2020pgw} yield two distinct mass squared differences: $\Delta m_{21}^2 = 7.506 \times 10^{-5}$ eV$^2$ and $\Delta m_{31}^2 = 2.55 \times 10^{-3}$ eV$^2$. Assuming that the lightest active neutrino is massless, we can deduce the following neutrino masses for active states: $m_{\nu_{a_i}} = (0.0, 0.0087, 0.0505)$ eV. Simultaneously, we aim to generate one sterile neutrino with masses of the order of $m_{\nu_{s_1}} \sim \mathcal{O}(10)$ keV and one with mass $m_{\nu_{s_{2}}} \sim \mathcal{O}(100)$ MeV to solve MiniBooNE excess events problem. One set of benchmark values capable of producing this intriguing sterile neutrino mass spectrum can be derived with mixing parameters on the order of $\mathcal{O}(10^{-4}-10^{-5})$~\cite{Dutta:2020scq}. 

The last three terms in Eq.~(\ref{eq:neutrinomass}) introduces interactions between the neutrinos and the scalars. Specifically, it generates interaction terms like $y_{\nu_i}\bar{\nu}_{a_i} \nu_{s_i} h_1$ and $y_{s_i}\bar{\nu}_{s_i} \nu_{s_i} h_1$. Because of neutrino mixing, the couplings $y_\nu$ and $y_s$ are proportional to $m_{{{1,2}n\psi}_i}$, which are unbounded and can therefore be treated as free parameters. Therefore, scalar couplings to neutrinos can be large compared to the other fermion couplings.   The light scalar field, $h_1$, would predominantly decay into $e^+e^-$ and $\bar{\nu}_{s_1} \nu_{s_1}$ pairs. 
We can adjust the couplings $y_{e{h_1}}$ and $y_{s_1}$ in such a way that we achieve approximately $Br(h_1 \rightarrow \bar{\nu}_{s_1} \nu_{s_1}) \simeq 0.95$ and $Br(h_1 \rightarrow e^+e^-) \simeq 0.05$, with the decay occurring promptly inside the detectors. For example, if we choose coupling $y_{s_1} \sim \mathcal{O}(10^-5)$ and $m_{\nu_{s_1}} \sim \mathcal{O}(10)$~keV, then the lifetime of $h_1$ becomes $10^{-14}$~sec.

\subsection{Gauge sector} \label{sec:gauge}

We denote the field strenth tensor of $U(1)_Y$ and $U(1)_H$ as $B_{\alpha \beta}$ and $X_{\alpha \beta}$ respectively. The kinetic energy can be expressed in the form, \begin{equation}
    \mathcal{L}_{kinetic} = -\frac{1}{4}B_{\alpha \beta}B^{\alpha \beta} -\frac{1}{4}X_{\alpha \beta}X^{\alpha \beta} -\frac{2 \alpha}{4}B_{\alpha \beta}X^{\alpha \beta}
\end{equation} where $\alpha$ is a physical parameter that gives the strength of kinetic mixing between the two Abelian fields and the requirement of positive kinetic energy requires that $|\alpha| <1$. After diagonalizing the kinetic term by a non-unitary transformation of the fields, the covariant derivative can be expressed in terms of the transformed fields as~\cite{Foot:1991kb}, 
\bea D_\rho \mathbb{I}_2 = \partial_\rho \mathbb{I}_2+ i \frac{g}{2} \tau^a W^a_\rho +i g^\prime Y B_\rho \mathbb{I}_2 +i \left( \frac{g_X}{2} Q_H + g^\prime_X \right) X_\rho \mathbb{I}_2  \eea where $g, g^\prime$~and~$g_X$ are the coupling constants of $SU(2)_L, U(1)_Y$~and~$U(1)_H$ gauge groups respectively with values $g= 0.64, g^\prime = 0.34$; $g^\prime_X = -g^\prime \alpha / \sqrt{1-\alpha^2}$ denotes the gauge coupling mixing between the two $U(1)$ groups;  $W^a_\rho, B_\rho$~and~$X_\rho$ are the corresponding gauge bosons in interaction basis; and $Y$ is the hypercharge and $Q_H$ is the charge under $U(1)_H$ gauge group. Now the effect of the kinetic mixing is embedded in the definition of the gauge covariant derivative.

Gauge boson masses and mixing can be calculated from the kinetic terms of the scalar fields. While $H_1$ is uncharged under the new gauge group, $|D_\mu H_1|^2$ contributes to the SM gauge boson masses and the new gauge boson mass term through the kinetic mixing effect. On the other hand, both $H_2$ and $\phi_H$ are charged under $U(1)_H$ and therefore contribute not only to the SM gauge boson masses but also to the new gauge boson mass directly. 

Therefore the terms of the Lagrangian giving gauge boson masses and mixing is ,\begin{equation}
    \mathcal{L} \subset \left( D^\rho H_1 \right)^\dagger \left( D_\rho H_1 \right) + \left( D^\rho H_2 \right)^\dagger \left( D_\rho H_2 \right) +\left( D^\rho \phi_H \right)^\dagger \left( D_\rho \phi_H \right)
\end{equation}

The charged gauge boson mass gets contribution from both $|D_\mu H_1|^2$ and $|D_\mu H_2|^2$ and is given by, $m_{W^{\pm}}^2 = g^2 v^2/4$. Now collecting all the mass terms for the neutral gauge bosons, the mass squared term can be written as, \be \label{eq:neutralgaugeint} \mathcal{L}^{0,gauge}_{mass}=\frac{1}{2}\left(B^\rho ~~~ W_3^\rho ~~~X^\rho  \right)~ \left(M^0_{gauge} \right)^2_{3\times3}~ \left( \begin{array}{c} B_\rho \\ W_{3\rho}  \\ X_\rho\end{array} \right) ~,~\, \ee where the mass square matrix is given by, \begin{equation}  \left(M^0_{gauge} \right)^2_{3\times3} = \frac{v^2}{4}\left( \begin{array}{ccc}{g^\prime}^2 &-g^\prime g &g^\prime  A_2 \\ -g^\prime g& g^2&-g  A_2 \\g^\prime  A_2 &-g  A_2 &  A_1\end{array} \right) ~.~\, \end{equation} where we have defined \bea A_2 &=& g^\prime_X + g_X A  \nonumber\\ A_1 &=& {g^\prime_X}^2+g_X^2 B +2 g_X g^\prime_X A \eea 

 where the quantities $A$ and $B$ are defined as, \bea A&=&\frac{v_2^2}{v_1^2 v_2^2} \\ B&=& \frac{v_2^2+v_\phi^2}{v_1^2 v_2^2} \eea

To obtain the masses of the physical gauge bosons, we aim to diagonalize the mass-squared matrix into the form $\left(M^0_{diag} \right)^2_{3\times3} = diag\left( m_A^2, m_Z^2, m_{Z^\prime}^2  \right)$. Note that $m_A^2=0$, which allows us to diagonalize this mass matrix using two block-diagonal rotations: one with Weinberg's angle $\theta_W$ and another with an additional angle $\xi$. This process can be expressed as follows:
\be \left(M^0_{diag} \right)^2_{3\times3} = diag\left( 0, m_Z^2, m_{Z^\prime}^2  \right) = R_{0,gauge}(\theta_W,\xi) \left(M^0_{gauge} \right)^2 R^T_{0,gauge}(\theta_W,\xi) \ee
where the diagonalization matrix is given by, \bea R_{0,gauge}(\theta_W,\xi) &=&  R_1(\xi) R_2(\theta_W) \nonumber\\ &=& \left( \begin{array}{ccc} 1 & 0 & 0 \\ 0 & \cos{\xi} & \sin{\xi} \\ 0 & -\sin{\xi} & \cos{\xi} \end{array} \right) \left( \begin{array}{ccc} \cos{\theta_W} & \sin{\theta_W} & 0 \\ -\sin{\theta_W} & \cos{\theta_W} & 0 \\ 0 & 0 & 1 \end{array} \right)   \eea

The mixing angles can be expressed in terms of  the gauge couplings: $\theta_W$, the Weinberg's angle, can be defined as $\tan{\theta_W} = g^\prime / g$; and $\xi$ is the new mixing angle, which can be expressed in terms of the new gauge coupling as, \begin{equation} \label{eq:xiangle}\tan{2\xi} = \frac{2 g_Z A_2}{A_1-g_Z^2}    \end{equation} where we have defined $g_Z = \sqrt{g^2 +{g^\prime}^2}$. The mass of the new gauge boson $Z^\prime$ depends on the choice of the couplings $g_X$ and $g_X^\prime$, as well as the vev $v_\phi$. It's worth noting that for $v_\phi = 10$ GeV, with couplings $g_X \sim \mathcal{O}(10^{-3})$ and $g_X^\prime \sim \mathcal{O}(10^{-4})$, we can generate a $Z^\prime$ with a mass of around $\mathcal{O}(10)$ MeV, which can explain the $e^+ e^-$ excess around 17 MeV reported by the ATOMKI Collaboration.

Note that we are primarily interested in the limit where $g_X$ and $g^\prime_X \ll 1$. In this limit, the expression for the angle can be simplified. First, note that $A$ and $B$ are of order $\mathcal{O}(1)$, so we have $A_1 - g_Z^2 \simeq -g_Z^2$. Therefore, the right-hand side of Eq.~(\ref{eq:xiangle}) is very small, allowing us to use the small angle approximation. In this limit, Eq.~(\ref{eq:xiangle}) can be written as \begin{equation} g_Z \xi = -g^\prime_X - g_X \sin^2{\beta}  \end{equation}

Define a vector $\left( Z^{\prime \rho}_{0} \right)^T \equiv \left( B^\rho, W^{3\rho}, X^\rho \right)$ in interaction basis and a vector $\left( Z^{ \rho}_{0} \right)^T \equiv \left( A^\rho, Z^{\rho}, Z^{\prime \rho} \right)$ in mass basis. Consequently, Eq.~(\ref{eq:neutralgaugeint}) can be written as, \bea \mathcal{L}^{0,gauge}_{mass} &=& \frac{1}{2} \left( Z^{\prime \rho}_{0} \right)^T R^T_{0,gauge}(\theta_W,\xi) \left(M^0_{diag} \right)^2 R_{0,gauge}(\theta_W,\xi) Z^{\prime }_{0 \rho} \nonumber\\ &=& \frac{1}{2} \left( Z^{ \rho}_{0} \right)^T \left(M^0_{diag} \right)^2 Z_{0 \rho} \eea where we have defined, \be Z_{0 \rho} = R_{0,gauge}(\theta_W,\xi) Z^{\prime }_{0 \rho}  \ee that connects the interaction states and the mass states.

Therefore, we can explicitly express the states in the interaction basis in terms of the mass basis as follows:  \begin{eqnarray} B_\mu &=& \cos{\theta_W} A_\mu -\sin{\theta_W} \cos{\xi} Z_\mu + \sin{\theta_W} \sin{\xi} Z^\prime_\mu ~,~\,\nonumber\\
W_{3\mu} &=&  \sin{\theta_W} A_\mu + \cos{\theta_W} \cos{\xi} Z_\mu  -\cos{\theta_W} \sin{\xi} Z^\prime_\mu ~,~\,\nonumber\\ 
X_\mu &=&   \sin{\xi} Z_\mu + \cos{\xi} Z^\prime_\mu~.~\,\end{eqnarray}

The terms of the model Lagrangian that give the interactions between the fermions and the gauge bosons are,
\begin{eqnarray} \label{eq:gauge-fermion}
    -\mathcal{L}_{kin,fermion} &=& \bar{q}^\prime_{L_j} i \gamma^\rho D_\rho {q}^\prime_{L_j} + \bar{u}^\prime_{R_j} i \gamma^\rho D_\rho u^\prime_{R_j} + \bar{d}^\prime_{R_j} i \gamma^\rho D_\rho d^\prime_{R_j}  \nonumber\\ &~& +  \bar{l}^\prime_{L_j} i \gamma^\rho D_\rho {l}^\prime_{L_j}  + \bar{n}^\prime_{R_j} i \gamma^\rho D_\rho n^\prime_{R_j} + \bar{e}^\prime_{R_j} i \gamma^\rho D_\rho e^\prime_{R_j} \nonumber\\ &~& + \bar{\psi}^\prime_L i \gamma^\rho D_\rho \psi^\prime_L + \bar{\psi}^\prime_R i \gamma^\rho D_\rho \psi^\prime_R
\end{eqnarray}

By expanding the terms in Eq.~(\ref{eq:gauge-fermion}), we obtain the  currents that couple to charged and neutral gauge bosons. It is important to note that the charged boson gauge current and the $A$ current remain the same as in the SM, while the $Z$ gauge current will be modified due to the neutral gauge boson mixing. In the limit of a small mixing angle, the SM $Z$ current can be recovered from the modified current. The most crucial interaction current for our analysis is the $Z^\prime$ current, which is given by:
\begin{equation}
    \mathcal{L}^{NC}_{Z^\prime}= - \sum_f \bar{f} \gamma^\rho \left( 
C_{f,V}+C_{f,A} \gamma^5 \right) f Z^\prime_\rho \end{equation} where we have defined, \begin{eqnarray}
    C_{f,V} &=& g_Z \sin{\xi} \left(-\frac{1}{2} T^3_f+\sin^2{\theta_W} Q_f \right) + g_X^\prime \cos{\xi} \left( Q_f-\frac{1}{2} T^3_f \right) \nonumber\\ 
    C_{f,A} &=& \frac{1}{2} \left( g_Z \sin{\xi} + g_X^\prime \cos{\xi}\right) T^3_f
    \end{eqnarray}
In the limit, $g_X, g_X^\prime \ll 1$ we can simplify the above expressions and get, \begin{eqnarray} \label{eq:gaugecoupling}
    C_{f,V} &=&- g_X \sin^2{\beta} \left(-\frac{1}{2} T^3_f+\sin^2{\theta_W} Q_f \right) + g_X^\prime \cos^2{\theta_W}  Q_f \nonumber\\ 
    C_{f,A} &=& -\frac{1}{2} g_X \sin^2{\beta} T^3_f
\end{eqnarray}
    
Note that the gauge interaction currents of $Z^\prime$ depend solely on the couplings $g_X$ and $g_X^\prime$. Additionally, the interactions of $Z^\prime$ with the physical mass states of neutrinos will be further suppressed by the mixing angles in the neutrino sector in comparison to the electron interactions. Therefore, we can neglect the neutrino interactions by appropriately choosing small mixing angles in the neutrino sector.

\section{Beryllium Anomaly at ATOMKI Experiment} \label{sec:ATOMKI}

In 2015, the ATOMKI spectrometer detected an anomalous peak at an opening angle of $140 \degree$, accompanied by a corresponding bump in the $(m_{ee})$ distribution during the IPC decays of $^8$Be$^* (18.15)$~\cite{Krasznahorkay:2015iga, Krasznahorkay:2018snd}. This behavior starkly contrasts with the expected SM behavior, where both the $(\theta_{ee})$ and $(m_{ee})$ distributions are anticipated to be monotonically falling curves. The bump observed in the data exhibited a remarkable statistical significance of $6.8 \sigma$ and could not be attributed to experimental or nuclear physics effects. Subsequently, in 2018, the experiment was repeated with an improved setup, reaffirming the presence of the anomaly and its consistency with the earlier measurements~\cite{Krasznahorkay:2018snd}. Additionally, it was discovered that the anomaly disappeared when the resonance peak was no longer present, leading to the conclusion that it originated from the decay of the resonant excited states. 

The observed bump in the data can be explained by the production of a massive, short-lived neutral particle with low velocity during the decay of $^8$Be$^* (18.15)$. This particle subsequently decays into an $e^+ e^-$ pair, leading to the emergence of a peak at a large opening angle in the experimental results. 
\begin{equation} ^8 \text{Be}^* (18.15) \rightarrow \text{$^8$Be}
+ X (X \rightarrow e^+e^-) 
\end{equation}  
The best-fit mass and decay branching ratio for this hypothetical neutral boson have been estimated with a goodness-of-fit statistic of $\chi^2/dof=1.07$~\cite{Krasznahorkay:2018snd}. 
\begin{eqnarray}
    &~& m_X = 16.70 \pm 0.35 (stat) \pm 0.5 (sys) ~\text{MeV} \\
    &~& \frac{\Gamma \left( ^8 \text{Be}^* (18.15) \rightarrow \text{$^8$Be}~ X \right)}{\Gamma \left( ^8 \text{Be}^* (18.15) \rightarrow \text{$^8$Be}~ \gamma \right)}Br (X \rightarrow e^+e^-) = 5.8 \times 10^{-6}
\end{eqnarray}

In 2019 and more recently in 2022, the ATOMKI collaboration successfully replicated its experiments  using $^4$He~\cite{Krasznahorkay:2019lyl, Krasznahorkay:2021joi} and $^{12}$C~\cite{Krasznahorkay:2022pxs} nuclei. They observed similar anomalous peaks, with $^4$He nuclei exhibiting an opening angle of $115 \degree$ and $^{12}$C nuclei showing peaks at $150\degree-160 \degree$. These observations are consistent with the presence of a new boson with a mass of 17 MeV. However, 
it is important to note that these results should be further confirmed and validated with additional statistics and repeated measurements. To establish the robustness of these findings and to test the hypothesis of the 17 MeV new boson, new and independent measurements are essential. One promising avenue is the use of MEG II at PSI~\cite{MEGII:2018kmf, Chiappini:2022egy} and the Montreal Tandem accelerator~\cite{Azuelos:2022nbu}, which have setups similar to that of the ATOMKI experiment. These facilities have the potential to replicate the measurements on $^8$Be nuclei and provide high-statistics data for a more comprehensive analysis. Given the high statistical  significance  and the potential for near future investigations, our primary focus remains on the Beryllium anomaly. Various attempts have been made to explain the ATOMKI anomaly by  new physics involving a 17 MeV bosonic field~\cite{Feng:2016jff, Feng:2016ysn, DelleRose:2017xil, DelleRose:2018eic, DelleRose:2018pgm, Nam:2019osu, Feng:2020mbt, Barducci:2022lqd, Denton:2023gat, Hostert:2023tkg}.

In the following, we demonstrate that the gauge interaction current of the new gauge boson $Z^\prime$ can explain the resonance bump in the Beryllium data while also satisfying all existing bounds. We start our analysis by writing the interactions of the fermions with the new gauge boson $Z^\prime$. The interactions can be written using Eq.~(\ref{eq:gaugecoupling}) as,
\begin{eqnarray} \label{eq:ZprimeATOMKI}
    -\mathcal{L}_{NC, Z^\prime} &=& \bar{u} \gamma^\rho \left[ \left\{ g_X \sin^2{\beta} \left( \frac{1}{4} - \frac{2}{3} \sin^2{\theta_W} \right)+\frac{2}{3} g_X^\prime \cos^2{\theta_W} \right\} + \left\{ -\frac{1}{4} g_X \sin^2{\beta} \right\} \gamma_5 \right] u Z^\prime_\rho \nonumber\\ &+&  \bar{d} \gamma^\rho  \left[ \left\{ g_X \sin^2{\beta} \left( -\frac{1}{4} + \frac{1}{3} \sin^2{\theta_W} \right)-\frac{1}{3} g_X^\prime \cos^2{\theta_W} \right\} + \left\{\frac{1}{4} g_X \sin^2{\beta} \right\} \gamma_5 \right] d Z^\prime_\rho \nonumber\\ &+& \bar{e} \gamma^\rho \left[ \left\{ g_X \sin^2{\beta} \left( -\frac{1}{4} +  \sin^2{\theta_W} \right)- g_X^\prime \cos^2{\theta_W}\right\} + \left\{ \frac{1}{4} g_X \sin{\beta}^2 \right\} \gamma_5 \right]  e Z^\prime_\rho
\end{eqnarray}

The interaction structure  of the $Z^\prime$ with fermions described in Eq.~\ref{eq:ZprimeATOMKI} is capable of resolving the Beryllium anomaly.  Before we identify the parameter space that matches the data, let's briefly outline a few important points,
\begin{enumerate}

    \item First, note that the strengths of the vector and axial-vector interactions are similar, i.e., $C_V^f \sim C_A^f$, for appropriate values of the couplings: $g_X \sim \mathcal{O}(10^{-3})$ and $g^\prime_X \sim \mathcal{O}(10^{-4})$. 

    \item For pure vector coupling, $\Gamma(^8\text{Be}^* \rightarrow ^8\text{Be} Z^\prime) \propto k_{Z^\prime}^3$, while for pure axial-vector coupling, $\Gamma(^8\text{Be}^* \rightarrow ^8\text{Be} Z^\prime) \propto k_{Z^\prime}$, where $k_{Z^\prime}$ is the momentum of the $Z^\prime$. Since we are considering a low-energy scenario, the momentum of the $Z^\prime$ is small. As we saw in the last point, $C_V^f \sim C_A^f$, therefore, we can neglect the effects of the vector coupling of $Z^\prime$ and its interference with the axial-vector component.

    \item A 17 MeV $Z^\prime$ can only decay to $e^+e^-$ and $\nu\bar{\nu}$.  The diphoton, $\gamma\gamma$, final state is prohibited by the Landau-Yang theorem~\cite{Landau:1948kw, Yang:1950rg}. A $3 \gamma$ final state is possible but highly suppressed. It's important to note that the interaction of the $Z^\prime$ with neutrinos will be suppressed by the neutrino mixing angle  compared to the electron couplings. Therefore, we can consider the coupling of the $Z^\prime$ to neutrinos as negligible. This helps circumvent the stringent constraints from neutrino-electron scattering experiments and ensures that the $Z^\prime$ mostly decays to an $e^+e^-$ pair with lifetime of the order of $\mathcal{O}(10^{-10})$~sec. 

\end{enumerate}

A light boson coupled to both quarks and leptons is subject to a wide range of experimental constraints. Below, we summarize the most significant constraints that would impact any potential explanation of the ATOMKI anomaly involving a BSM degree of freedom with a mass of approximately 17 MeV in the case of a spin-1 particle.
\begin{enumerate}
    \item {\textbf {Rare neutral pion decay}}: The constraints on the quarks' coupling to $Z^\prime$ arise from the study of the rare pion decay process, $\pi^0 \rightarrow \gamma Z^\prime$( $Z^\prime \rightarrow e^+ e^-$). Currently, the NA48/2 experiment~\cite{NA482:2015wmo, Raggi:2015noa} provides the strongest bounds on this interaction. These constraints are proportional to the vector-vector anomaly factor $|C_{u,V} Q_u - C_{d,V} Q_d|^2$, which in turn depends on the vector couplings of the $Z^\prime$. The bound can be expressed as:$|2 C_{u,V} +C_{d,V}| \lesssim 2.4 \times 10^{-4} / \sqrt{\text{Br}(Z^\prime \rightarrow e^+ e^-)} $ for $m_{Z^\prime} \simeq 17$~MeV~\cite{Feng:2016jff, Feng:2016ysn}. It's worth noting that the contribution of the axial vector components is suppressed due to the light quark masses, as these components are induced by chiral symmetry breaking. 

    \item  {\textbf {Anomalous magnetic moment of charged leptons}}: The universal couplings of the $Z^\prime$ boson to charged leptons affect both electron and muon anomalous magnetic moments. As the $Z^\prime$ decays into $e^+ e^-$, it possesses couplings to both electron and muon, leading to new contributions to their magnetic moments. Due to the $Z^\prime$ boson's vector and axial-vector couplings to charged leptons, the contribution to the magnetic moment is negative for a $Z^\prime$ mass around 17 MeV.  However, in our model, a positive contribution arises from the light scalar particle, which has non-universal couplings to charged leptons. By balancing these positive and negative contributions, correct values for the anomalous magnetic moments can be achieved. See section.~\ref{sec:g2} for details.

     \item {\textbf{Electron-positron scattering}}: An upper limit on the coupling of the $Z^\prime$ boson to electrons can be determined through the investigation of $Z^\prime$ production associated with initial-state radiation in $e^+ e^-$ collisions, specifically the process $e^+ e^- \rightarrow \gamma Z^\prime$, where the $Z^\prime$ subsequently decays into $e^+ e^-$. KLOE-2 has provided an upper bound on this coupling~\cite{Anastasi:2015qla}, given by $\sqrt{C_{e,V}^2 + C_{e,A}^2} \lesssim 6 \times 10^{-4}/ \sqrt{\text{Br}(Z^\prime \rightarrow e^+ e^-)}$. 
    
    \item {\textbf{Electron beam dump experiments}}: A lower limit on the coupling of the $Z^\prime$ boson to electrons can be derived from fixed-target electron beam dump experiments such as NA64~\cite{NA64:2018lsq} and E141~\cite{Riordan:1987aw}. These experiments investigate the bremsstrahlung interaction of electrons with target nuclei, leading to the process $e+N(A,Z) \rightarrow e+N(A,Z)+Z^\prime$, followed by the subsequent decay of the newly produced gauge boson, $Z^\prime \rightarrow e^+ e^-$. In both cases, these experiments search for decays-in-flight of the $Z^\prime$. The absence of any such direct detection results in bounds, which are given by $\sqrt{C_{e,V}^2 + C_{e,A}^2} \gtrsim 3.6 \times 10^{-5} / \sqrt{\text{Br}(Z^\prime \rightarrow e^+ e^-)}$ for NA64~\cite{NA64:2019auh, Barducci:2022lqd} and $\sqrt{C_{e,V}^2 + C_{e,A}^2} \gtrsim 6 \times 10^{-5} / \sqrt{\text{Br}(Z^\prime \rightarrow e^+ e^-)}$ for E141~\cite{Andreas:2012mt}, specifically for $m_{Z^\prime} \simeq 17$ MeV.

    \item {\textbf{Møller scattering}}: The SLAC E158 experiment~\cite{SLACE158:2005uay} conducted measurements of the parity-violating asymmetry in Møller scattering, a process that is sensitive to the existence of a Z' boson. Through a comparison between the experimentally measured asymmetry and theoretical predictions, limitations can be imposed on the product $C_{e,V} C_{e,A}$ for various Z' boson masses. In the case of $m_{Z^\prime} \simeq 17$ MeV, the experiment establishes a constraint of $|C_{e,V} C_{e,A}| \lesssim 10^{-8}$~\cite{Kahn:2016vjr}. 

    \item {\textbf{Atomic parity violation}}:The measurement of the weak nuclear charge, denoted as $Q_W^{\text{eff}}(Z,N)$ in the cesium (Cs) atom, imposes the most stringent constraints in terms of atomic parity violation (APV)~\cite{Davoudiasl:2012ag, Bouchiat:2004sp}. This measurement necessitates that the contribution stemming from new physics to the weak charge must satisfy $|\Delta Q_W|\lesssim 0.71$ at the $2\sigma$ confidence level~\cite{Porsev:2009pr}. Here, $\Delta Q_W(Z,N) = Q_W^{\text{eff}}(Z,N)-Q_W^{\text{SM}}(Z,N)$ represents the difference in weak charge between the actual measurement and the Standard Model prediction. This bound translates into a constraint on the product of $C_{e,A}$ and a combination of $C_{u,V}$ and $C_{d,V}$ based on the definition of $\Delta Q_W$. Specifically, it is given by the equation: \begin{equation}      \Delta Q_W(Z,N) = - \frac{2\sqrt{2}}{G_F} \frac{C_{e,A}}{m_{Z^\prime}^2} \left[ (2 C_{u,V}+C_{d,V})Z+(C_{u,V}+2C_{d,V})N \right] . \end{equation}

    \item {\textbf{Neutrino based experiments}}: Various experiments involving neutrinos in either the initial and final states, such as neutrino-electron scattering~\cite{Denton:2023gat}, coherent elastic neutrino-nucleus scattering (CE$\nu$NS)~\cite{Denton:2023gat}, and decays of charged pions~\cite{Hostert:2023tkg}, kaons~\cite{Davoudiasl:2014kua}, can impose constraints on combinations of couplings, including those involving $C_{\nu}$. Since we assume very small neutrino mixing angles, $\sim \mathcal{O}(10^{-6})$  and, consequently, small neutrino couplings, our model automatically satisfies all of these constraints.
    \end{enumerate}

Therefore, the primary constraints on the $Z^\prime$ couplings, under the assumption that the $Z^\prime$ boson primarily decays to an $e^+e^-$ pair, can be summarized as follows.  \begin{eqnarray} \label{eq:atomkibounds}  |2 C_{u,V}+C_{d,V}| &~&\lesssim ~ 2.4 \times 10^{-4};  \nonumber\\    3.6 \times 10^{-5} ~\lesssim ~\sqrt{C_{e,V}^2 +C_{e,A}^2} &~&\lesssim~ 6 \times 10^{-4}; \nonumber\\    |C_{e,V} C_{e,A}| &~&\lesssim ~10^{-8}; \nonumber\\     |\Delta Q_{W}(Cs)| &~&\lesssim ~ 0.7; \end{eqnarray}

\begin{figure}[tbp]	
\centering
\begin{subfigure}{0.49\textwidth}\includegraphics[width=0.97\linewidth]{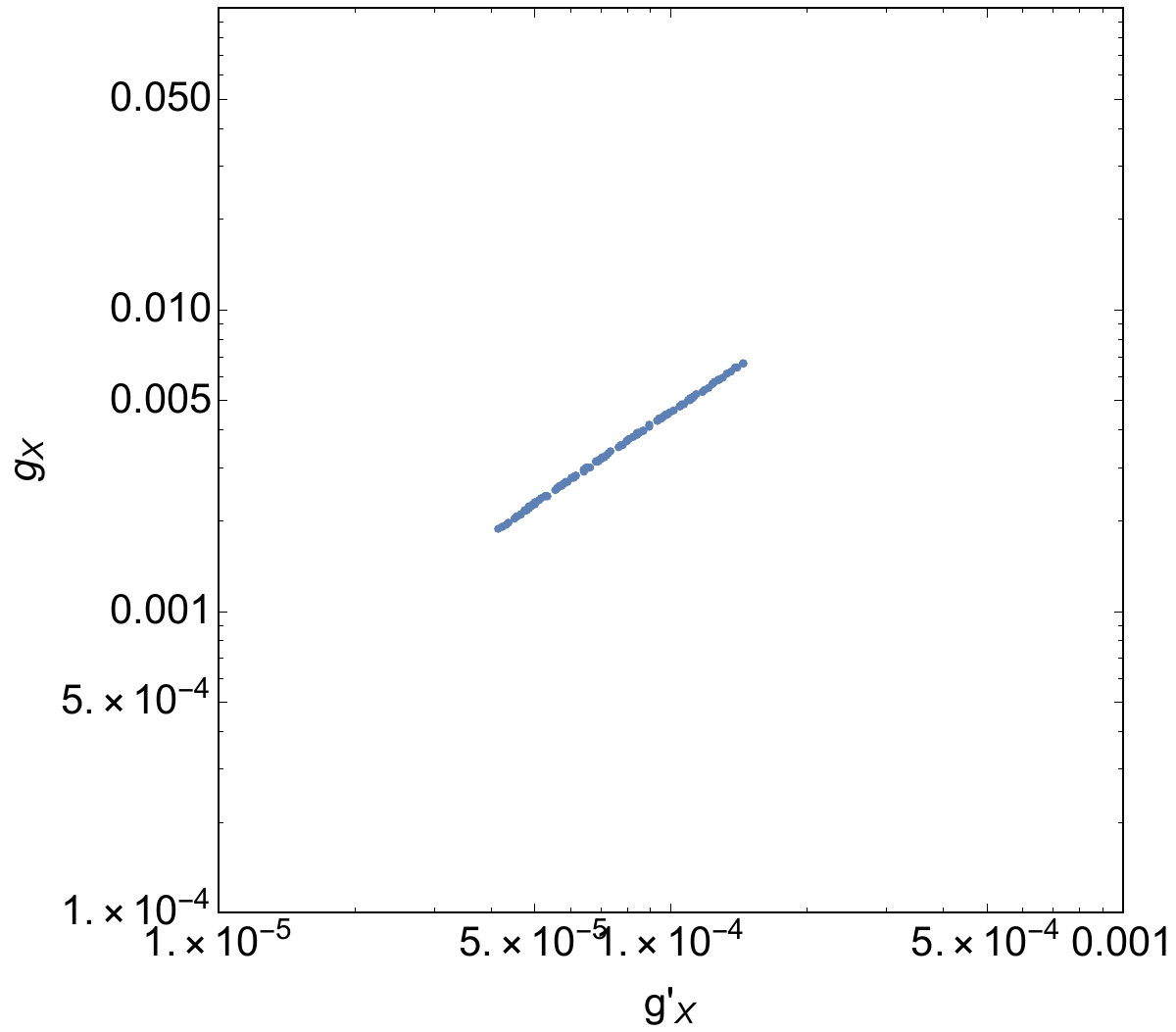}
\caption{\label{fig:gXpvsgX}}							\end{subfigure}	
\begin{subfigure}{0.49\textwidth}		\includegraphics[width=0.97\linewidth]{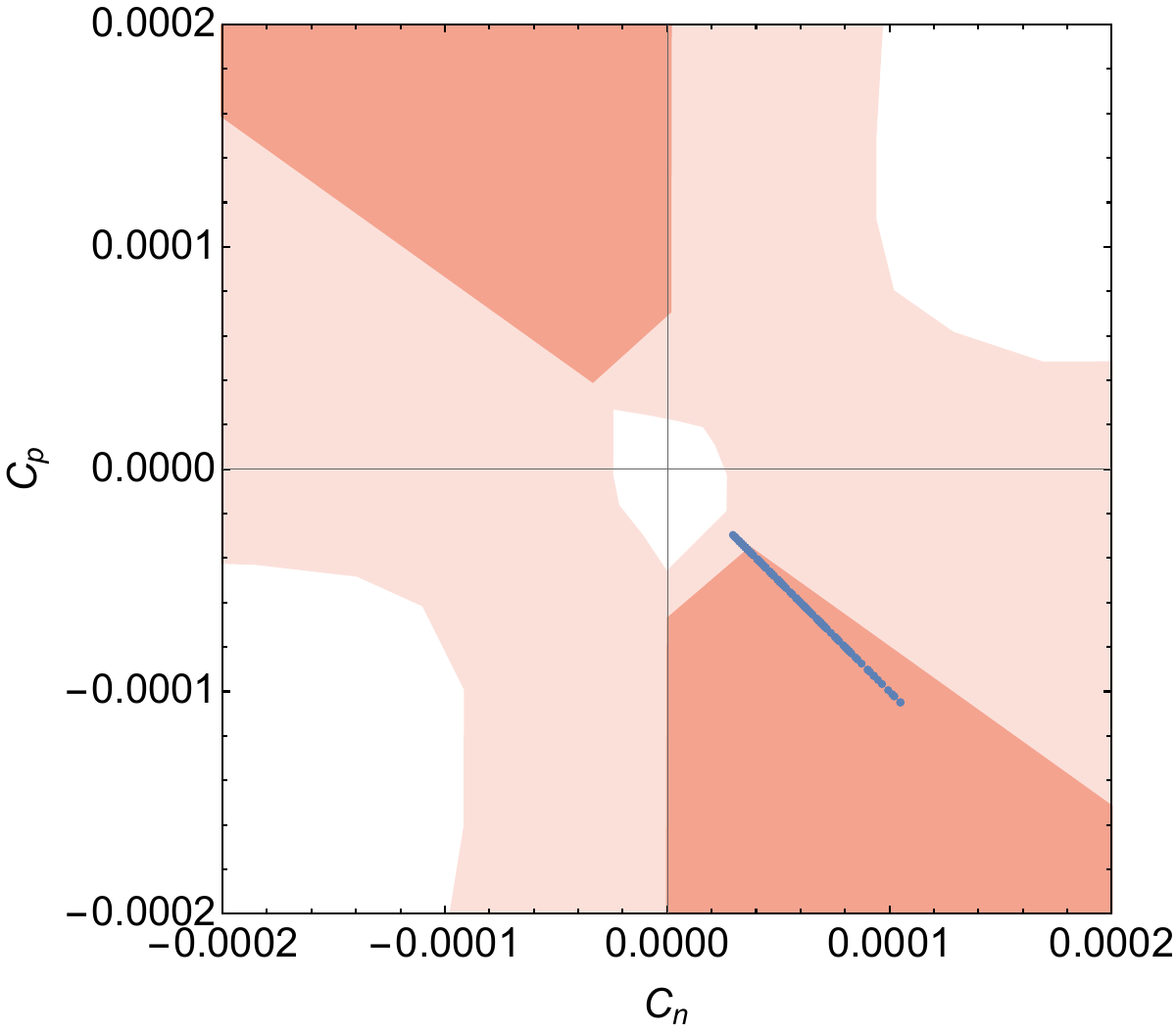}				
\caption{\label{fig:ATOMKIfit}}							\end{subfigure}	
	
\captionsetup{justification   = RaggedRight,
             labelfont = bf}
\caption{\label{fig:ATOMKI} (a) The allowed parameter range in the $(g_X,g_X^\prime)$ plane that satisfies all the  relevant constraints summarized in Eq.~(\ref{eq:atomkibounds}).(b)The blue points represent the values of axial coupling to protons/neutrons, denoted as $C_{p/n}$, obtained using the approved range of $(g_X, g_X^\prime)$ from the preceding figure. In this context, the dark (light) shaded area corresponds to the $1\sigma$ ($2\sigma$) acceptable region for $C_{p/n}$ in the case of a general axial vector coupling reproduced from Ref.~\cite{Barducci:2022lqd}. }
\end{figure}

All these constraints can be projected into the $(g_X, g_X^\prime)$ plane, as all the coupling coefficients depend on the gauge coupling of the $U(1)_H$ gauge group, denoted as $g_X$, and the gauge coupling mixing parameter, $g_X^\prime$. The allowed parameter region in the $(g_X, g_X^\prime)$ plane that satisfies all the above constraints is depicted in figure.~\ref{fig:gXpvsgX}. Note that the couplings $g_X$ and $g_X^\prime$ are linearly dependent on each other within the allowed region. We find that the allowed range of values for the parameter $g_X$ is $(1.8-6.6)\times 10^{-3}$, and for $g_X^\prime$, it is $(0.4-1.4) \times 10^{-4}$.

To perform the nuclear decay calculation, one must define the coupling of $Z^\prime$ to nucleons. When considering only the axial vector coupling, it can be expressed as: 
\begin{equation}    
\mathcal{L}_{Z^\prime}^N = C_p \bar{p} \gamma^\rho \gamma^5 p Z^\prime_\rho + C_n \bar{n} \gamma^\rho \gamma^5 n Z^\prime_\rho 
\end{equation} 

where couplings $C_{p/n}$ can be represented in terms of the axial vector couplings of up and down type quarks as follows: 
\begin{eqnarray} 
C_p &=& g_{u,A} C_{u,A}+ g_{d,A} C_{d,A}; \nonumber\\  
C_n &=& g_{d,A} C_{u,A}+ g_{u,A} C_{d,A}; 
\end{eqnarray} 
where the coefficients $g_{u/d,A}$ can be obtained from the results of neutron $\beta$-decay~\cite{Alexandrou:2021wzv} and Lattice QCD calculations~\cite{Alexandrou:2019brg} and are given as follows: $g_{u,A}= 0.817$ and $g_{d,A}=- 0.450$. It's worth noting that $C_{u,A} = -C_{d,A}\equiv C$. Consequently, a simple relationship between $C_p$ and $C_n$ can be established: $C_p = -C_n = 1.267 C$ where $C$  depends on $g_X$ only. By employing the allowable values of $g_X$  from figure.~\ref{fig:gXpvsgX}, we can determine the corresponding values of $C_p$ and $C_n$ , which are presented in figure.~\ref{fig:ATOMKIfit} alongside the $1 \sigma$ (dark shaded region) and $2 \sigma$ (light shaded region) allowed regions of effective nuclear couplings. In particular, the values of $C_p$ are within the range of $-(0.297 - 1.050)\times 10^{-4}$, and $C_n$ falls within the range of $(0.297 - 1.050)\times 10^{-4}$. These values have the potential to explain the excess observed in the  $^8$Be nuclear decay at the ATOMKI experiment. Note that most of the values are within the $1 \sigma$ allowed region. Also note that, while the axial couplings $C_{p/n}$ depend solely on $g_X$, the coupling $g_X$ itself are related to $g_X^\prime$ through the constraint conditions in Eq.~(\ref{eq:atomkibounds}).

\section{Muon Anomalous Magnetic Moments} \label{sec:g2}

A persistent and significant discrepancy between experimental data and the theoretical predictions of the SM lies in the anomalous magnetic moment of the muon, denoted as $a_{\mu} = (g_\mu -2)/2$.  Notably, the E989 experiments at FNAL have recently released the most precise measurement of $a_{\mu}$~\cite{Muong-2:2023cdq}, achieved through the combination of data from Runs 1, 2, and 3. This refined value is 
\begin{equation}  
a_{\mu}^{FNAL}= 116 ~592~055(24) \times 10^{-11},  
\end{equation} upon combining this result with those from BNL
~\cite{Muong-2:2006rrc}, gives a comprehensive experimental average 
\begin{equation}  a_{\mu}^{exp}= 116 ~592~059(22) \times 10^{-11}.
\end{equation}

The FNAL result is entirely consistent with the most accurate previous measurements~\cite{Abi:2021gix} and, notably, enhances precision by a factor of two. Additional analysis of the remaining data is anticipated to yield a further twofold improvement in statistical accuracy. Simultaneously, the upcoming E34 experiment at J-PARC~\cite{Saito:2012zz, Lee:2019xkg} is dedicated to minimizing statistical uncertainties and independently confirming the FNAL findings.

On the theoretical front, a global collaboration known as the Muon $g-2$  Theory Initiative has offered the most precise and comprehensive predictions for the SM value~\cite{Aoyama:2020ynm}, which is calculated to be 
\begin{equation}  
a_{\mu}^{th}= 116 ~591~810(43) \times 10^{-11}.
\end{equation}
This SM prediction relies on the most contemporary and precise calculations encompassing Quantum Electrodynamics (QED)~\cite{Aoyama:2012wk, Aoyama:2019ryr}, Electroweak interactions~\cite{Czarnecki:2002nt, Gnendiger:2013pva}, Hadronic Vacuum Polarization (HVP)~\cite{Davier:2017zfy, Keshavarzi:2018mgv, Colangelo:2018mtw, Hoferichter:2019mqg, Davier:2019can, Keshavarzi:2019abf, Kurz:2014wya}, and Hadronic Light-by-Light Scattering~\cite{Melnikov:2003xd, Masjuan:2017tvw, Colangelo:2017fiz, Hoferichter:2018kwz, Gerardin:2019vio, Bijnens:2019ghy, Colangelo:2019uex, Pauk:2014rta, Danilkin:2016hnh, Jegerlehner:2017gek, Knecht:2018sci, Eichmann:2019bqf, Roig:2019reh, Colangelo:2014qya, Blum:2019ugy}. Specifically, the HVP, which constitutes the leading-order hadronic contribution, was derived from measurements of the electron-positron to hadrons cross-section performed by various experiments. However, the BMW collaboration recently introduced improvements in the theoretical prediction for HVP through lattice QCD calculations, introducing a noticeable tension with other data~\cite{Borsanyi:2020mff}. Furthermore, preliminary measurements of the electron-positron to $\pi ^+ \pi^-$ cross-section from the CMD-3 experiment~\cite{CMD-3:2023alj} have exhibited substantial tension. It's important to note that these results are awaiting independent verification, and multiple ongoing theoretical endeavors aim to clarify the current state of theoretical understanding~\cite{Colangelo:2022jxc}.

\begin{figure}[tbp]	
\centering
\begin{subfigure}{0.49\textwidth}\includegraphics[width=0.97\linewidth]{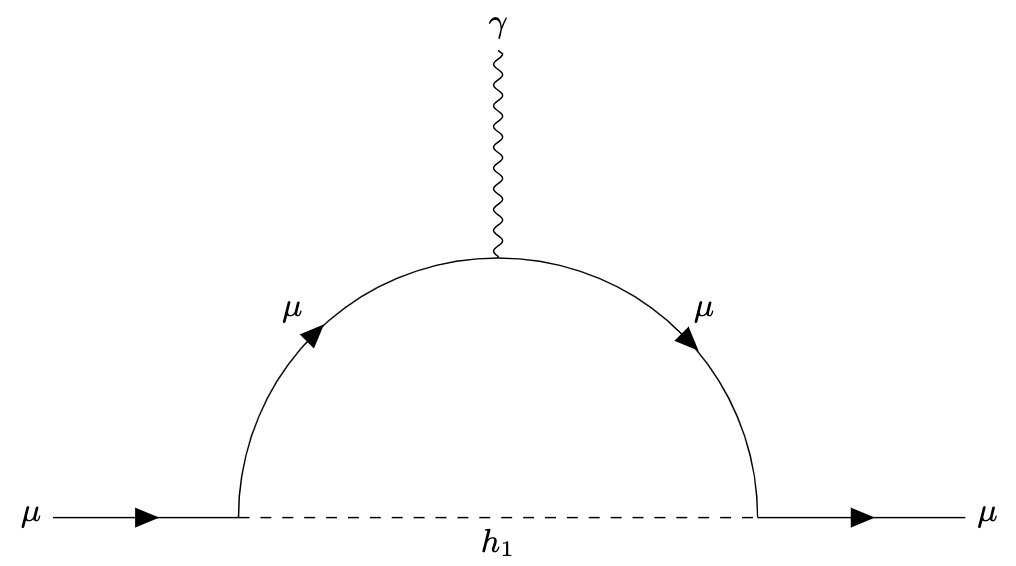}
\caption{\label{fig:g2h1}}							\end{subfigure}	
\begin{subfigure}{0.49\textwidth}		\includegraphics[width=0.97\linewidth]{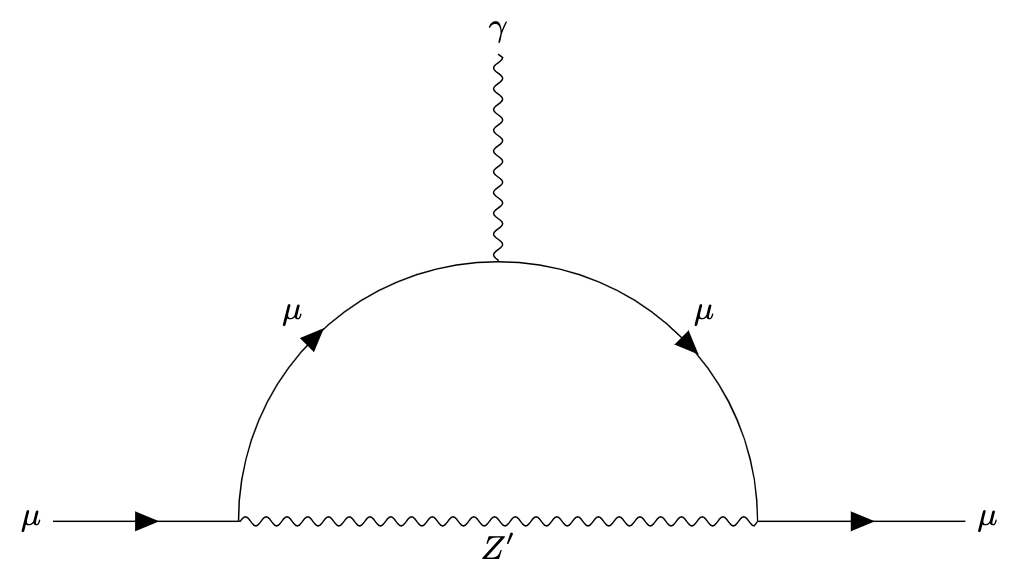}				
\caption{\label{fig:g2Zp}}							\end{subfigure}	
	
\captionsetup{justification   = RaggedRight,
             labelfont = bf}
\caption{\label{fig:g2fig} One-loop diagrams mediated by the light scalar $h_1$ and the light vector $Z^\prime$ that contribute to the anomalous magnetic moments of muon.}
\end{figure}
Hence, when we compare the experimental and theoretical values, a discrepancy of 5.1 sigma emerges,
\begin{equation} 
\Delta a_\mu =a_{\mu}^{exp}-a_{\mu}^{th}= 2.49 (0.48) \times 10^{-9},
\end{equation} 
with the uncertainties being combined by adding them in quadrature. 
Within the framework of our model, the muon's magnetic moment receives  contributions from both the light scalar $h_1$ and the light vector $Z^\prime$ mediated through one-loop diagrams. 
The relevant diagrams are shown in figure.~\ref{fig:g2fig}.

The Lagrangian in the physical mass basis, which leads to the expression for the magnetic moment of muon depicted in figure.~\ref{fig:g2fig}, can be written as follows: \begin{equation}
    -\mathcal{L} \subset y_{\mu h_1} \bar{\mu}  \mu h_1 +  \bar{\mu} \gamma^\rho \left( C_{\mu,V}+ C_{\mu,A} \gamma^5 \right)  \mu Z^\prime_\rho
\end{equation}

We can derive the contribution to  $\Delta a_\mu$ from figure.~\ref{fig:g2h1} as follows: 
\begin{equation}    
\Delta a_{\mu}^{(h_1)} = y_{\mu h_1}^2\frac{m_\mu^2}{8 \pi^2 } 
\int_0^1 dx \frac{2 x^2 -x^3}{m_\mu^2 x^2 + m_{h_1}^2 (1-x)} 
\end{equation} 
Similarly the  vector boson contribution from figure.~\ref{fig:g2Zp} is: \begin{equation}  
\Delta a_{\mu}^{(Z^\prime)} = \frac{m_\mu^2}{4 \pi^2} \int_0^1 dx \frac{\left[ C_{\mu, V}^2 x^2 (1-x) + C_{\mu, A}^2 \{ x(1-x)(x-4) - \frac{4 m_\mu^2 x^3}{m_{Z^\prime}^2} \} \right]}{m_\mu^2 x^2 + m_{Z^\prime}^2 (1-x)} 
\end{equation} 
where
\bea 
C_{\mu,V} &=& g_X \sin^2{\beta} \left( -\frac{1}{4} +  \sin^2{\theta_W} \right)- g_X^\prime \cos^2{\theta_W} \nonumber\\ C_{\mu,A}&=& \frac{1}{4} g_X \sin{\beta}^2
\eea
Hence, the overall contribution to the anomalous magnetic moment is expressed as: 
\begin{equation}  
\Delta a_\mu =  \Delta a_{\mu}^{(h_1)} + \Delta a_{\mu}^{(Z^\prime)} 
\end{equation}

 For a mediator mass around $\mathcal{O}(10)$~MeV, the $\Delta a_{\mu}^{(Z^\prime)}$ contribution consistently appears negative. Conversely, $\Delta a_{\mu}^{(h_1)}$ always yield a positive contribution. To achieve the correct value for $\Delta a_\mu$, these contributions are balanced against each other. A similar approach was taken in the case of low scale $U(1)_{T3R}$ theory as well~\cite{Dutta:2021afo}. Note that $C_{\mu,V}$ and $C_{\mu, A}$ rely on the gauge coupling of the $U(1)_H$ gauge group denoted as $g_X$ and the gauge coupling mixing parameter $g_X^\prime$.  The relationship between $g_X$ and $g_X^\prime$ has been already determined by addressing the ATOMKI anomaly in Sec.~\ref{sec:ATOMKI}, given in figure.~\ref{fig:gXpvsgX}. The scalar coupling to muon is bounded by muon mass. We examine four distinct benchmark scenarios, each featuring a different light scalar mass, to calculate the total contribution to $\Delta a_\mu$. Throughout these scenarios, we maintain a fixed $Z^\prime$ mass of 17 MeV.   A summary of the values is provided in Table.~\ref{table:g2fit}. The primary constraints on the $(m_{h_1}, y_{\mu h_1})$ plane arise from fixed-target/beam dump experiments like E137~\cite{Dobrich:2015jyk, Dolan:2017osp, Bjorken:1988as, Batell:2016ove} and Orsay~\cite{Batell:2016ove}. We determine that couplings of the order of $\mathcal{O}(10^{-4}-10^{-3})$ are permissible for $m_{h_1} \sim \mathcal{O}(10-100)$ MeV~\cite{Dutta:2020scq}. Future experiment such as FASER~\cite{Feng:2017uoz, Feng:2017vli, Batell:2017kty}, ShiP~\cite{Alekhin:2015byh, Batell:2017kty}, and Fermilab $\mu-$beam fixed target~\cite{Chen:2017awl, Batell:2017kty} can probe this paramater space.

 \begin{table}[tbp]
     \centering
     \begin{tabular}{ccccc}
     \hline
         $m_{h_1}$ & $g_X$ & $g_X^\prime$ & $y_{\mu h_1}$ & $\Delta a_\mu$\\
         &&&& \\ 
         \hline
         12 MeV & $1.8 \times 10^{-3}$ & $0.4\times 10^{-4}$ & $0.4\times 10^{-3}$ & $1.98 \times 10^{-9}$\\
         15 MeV & $1.8 \times 10^{-3}$ & $0.4\times 10^{-4}$ & $0.4\times 10^{-3}$ & $1.89 \times 10^{-9}$\\
         20 MeV & $1.8 \times 10^{-3}$ & $0.4\times 10^{-4}$ & $0.4\times 10^{-3}$ & $1.78 \times 10^{-9}$\\
         50 MeV & $1.6 \times 10^{-3}$ & $0.39\times 10^{-4}$ & $0.4\times 10^{-3}$ & $1.08 \times 10^{-9}$\\
         \hline
     \end{tabular}
     \captionsetup{justification   = RaggedRight,
             labelfont = bf}
     \caption{ \label{table:g2fit} We present four benchmark scenarios, each with a different light scalar mass, to obtain values of $\Delta a_\mu$. Here, we have kept the vector boson mass fixed at $m_{Z^\prime} = 17$ MeV.}
    
 \end{table}

One-loop diagrams, similar to those illustrated in figure.~\ref{fig:g2fig} but involving electrons as the charged leptons, can impact the magnetic moment of the electron. Despite both the muon and electron having the same coupling strength to the vector boson $Z^\prime$, their Yukawa couplings differ. Specifically, the electron Yukawa coupling, denoted as $y_{e h_1}$, is of the order of $\mathcal{O}(10^{-6})$ in our model. However, it's essential to note that the contribution from the scalar-mediated diagram is three orders of magnitude smaller than that from the vector-mediated diagram within the relevant parameter space. Consequently, the total contribution to the electron's magnetic moment amounts to approximately $\mathcal{O}(10^{-14})$, which is one order of magnitude smaller than the predicted value and predominantly comes from the $Z^\prime$ contribution.

\section{MiniBooNE Excess Events} \label{sec:MB}

In 2020, after collecting data for 17 years from 2002 to 2019, MiniBooNE updated its results~\cite{Aguilar-Arevalo:2020nvw}. They reported an excess of single shower ($1-sh$) events, with $560.6\pm 119.6$ events in the neutrino mode and $77.4 \pm 28.5$ events in the antineutrino mode, over the estimated background within the energy range of 200 to 1250 MeV. These observations were based on $11.27 \times 10^{20} $protons-on-target (POT) in the antineutrino mode and $18.75 \times 10^{20}$ POT in the neutrino mode. The combined excess events totaled $638.0 \pm 132.8$, leading to a $4.8 \sigma$ tension with the predictions of the two-neutrino oscillation scenario within the Standard Model.

The initial explanation for the excess observed in MiniBooNE data was the $3+1$ oscillation scenario~\cite{Sorel:2003hf, Karagiorgi:2009nb, Giunti:2011cp, Kopp:2011qd, Abazajian:2012ys, Conrad:2012qt, Kopp:2013vaa, Collin:2016aqd, Gariazzo:2017fdh, Asaadi:2017bhx, Dentler:2018sju, Boser:2019rta, Diaz:2019fwt}, which introduced an eV-scale sterile neutrino responsible for inducing short-distance oscillations: $\nu_\mu \rightarrow \nu_e$. However, this hypothesis encounters tension with other neutrino data~\cite{Collin:2016rao, MINOS:2020iqj, IceCube:2020phf, IceCube:2020tka} and conflicts with cosmological observations~\cite{Hamann:2011ge, Archidiacono:2013xxa, Hagstotz:2020ukm}. Recent findings from the MicroBooNE experiment~\cite{MicroBooNE:2016pwy} at Fermilab (FNAL) have added complexity to the situation. MicroBooNE results have cast doubt on the notion that neutral current $\Delta \rightarrow N \gamma$ backgrounds alone are responsible for the observed excess, with a confidence level of $94.8\%$ in disfavor of this explanation~\cite{MicroBooNE:2021zai}. Furthermore, the idea that generic $\nu_e$ interactions are the primary cause of the excess is also challenged by MicroBooNE data~\cite{MicroBooNE:2021tya}. These observations strongly advocate the necessity for new physics to explain the excess. It's worth noting that MiniBooNE cannot distinguish between photons and electrons, whereas MicroBooNE's advanced calorimetry and precise resolution enable a clear separation between electrons and photons~\cite{Contin:2017mck, MicroBooNE:2021nxr}. This capability allows MicroBooNE to explore BSM physics as a potential solution to this low-energy excess.

Numerous new physics scenarios have been proposed to account for this anomaly. Among them, solutions based on neutrinos~\cite{Bertuzzo:2018itn, Ballett:2018ynz, Ballett:2019cqp, Ballett:2019pyw, Fischer:2019fbw, Dentler:2019dhz, deGouvea:2019qre, Abdallah:2020biq, Datta:2020auq, Dutta:2020scq, Abdullahi:2020nyr, Abdallah:2020vgg, Abdullahi:2023ejc} hold particular promise, as they can explain the absence of excess events in the MiniBooNE beam dump mode~\cite{MiniBooNEDM:2018cxm}. Within the realm of neutrino-based solutions, several scenarios are conceivable~\cite{Brdar:2020tle}. One particularly intriguing possibility involves the coherent upscattering of light neutrinos to heavy neutrinos, followed by the decay of the heavy neutrino into an electron-positron pair through the mediation of a light particle~\cite{Bertuzzo:2018itn, Dutta:2020scq}. The neutrino's energy can be reconstructed by analyzing the energy and angular distributions of the mediator arising from the heavy neutrino's decay~\cite{Martini:2012fa}. However, many of these neutrino-based solutions, especially those involving vector mediators, face potential conflicts with the absence of signal observations in other neutrino experiments like CHARM-II~\cite{CHARM-II:1989nic, CHARM-II:1992vao, CHARM-II:1994dzw, Arguelles:2018mtc}, MINER$\nu$A~\cite{MINERvA:2015jih, MINERvA:2015nqi, MINERvA:2019hhc}, and T2K ND280~\cite{Kudenko:2008ia, T2K:2011qtm, Assylbekov:2011sh, T2KND280FGD:2012umz, T2K:2012bge, T2K:2019jwa, T2K:2020lrr}. This discrepancy arises because the coherent scattering cross-section for the vector mediator becomes more pronounced with increasing neutrino energy~\cite{Arguelles:2018mtc}. Conversely, when mediated by a scalar particle, the cross-section diminishes with rising neutrino energy and therefore does not lead to any excess events in other neutrino experiments~\cite{Dutta:2020scq}.

\begin{figure}[tbp]
\centering
\includegraphics[height=12cm,width=14cm]{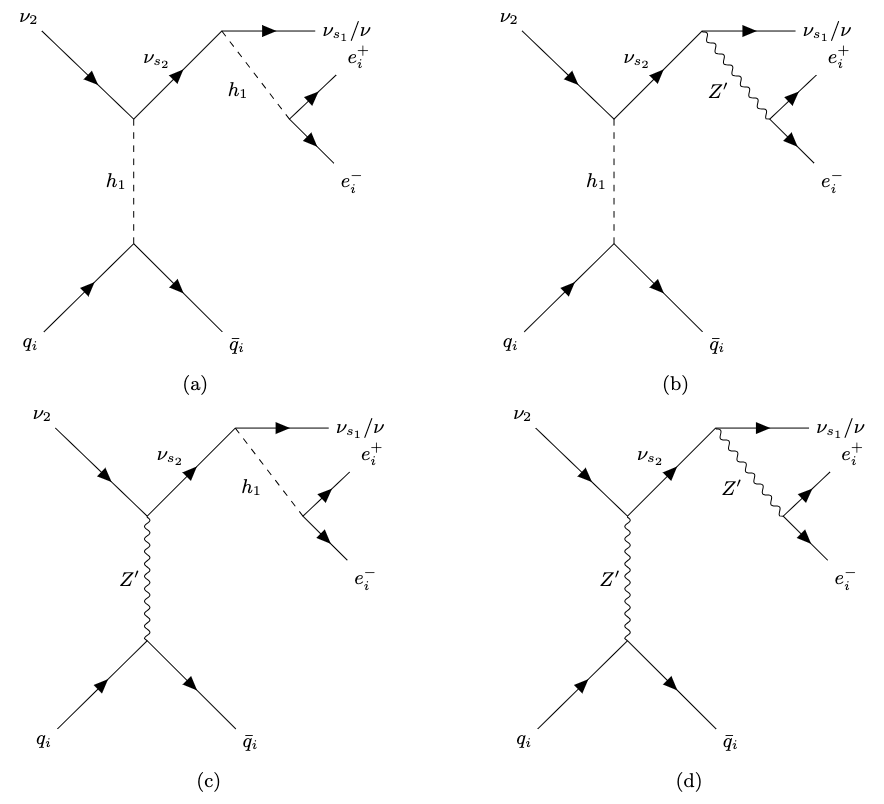}
\captionsetup{justification   = RaggedRight,
             labelfont = bf}
\caption{\label{fig:MB} The Feynman diagrams represent the upscattering-double decay scenario, involving both light mediators, $h_1$ and $Z^\prime$, contributing to the MiniBooNE excess events in our model.
 }
\end{figure}
In the context of our model, the heavy sterile neutrino, denoted as $\nu_{s_2}$, is generated through the upscattering process: $\nu_\mu A \rightarrow \nu_{s_2} A$. Importantly, in our model, this upscattering process can be mediated by both of the two light mediators, 17 MeV $Z'$ for the ATOMKI anomaly and light $h_1$ for the muon $g-2$. Note that, since this upscattering process is coherent, it experiences an enhancement by a factor of $A^2$. Subsequently, the produced heavy neutrino promptly undergoes decay, leading to either the lightest sterile neutrino or active neutrinos, along with an on-shell mediator, either $Z'$ of $h_1$. This mediator, in turn, decays into a pair of $e^+e^-$. This specific scenario can be categorized as an upscattering-double decay scenario, as identified in Ref.~\cite{Brdar:2020tle}. In this scenario, there are four possible diagrams, as illustrated in figure.~\ref{fig:MB}, that can contribute to the observed excess of events. Our plan is 
\begin{enumerate}
    \item First, we perform calculations for the contribution of all four diagrams individually and then sum their results. Let's denote the predicted number of events from our model as $N_{th}$, and the corresponding experimental value as $N_{exp}$. In this context, we anticipate that $N_{th}^{MB} = N_{exp}^{MB} $, within some permissible deviation. Here $MB$ stands for MiniBooNE. 

    \item Next, our objective is to evaluate whether the parameter space of this model yields ambiguous predictions for the number of events in other neutrino experiments. We can express the predicted number of events in these experiments as $N_{exp}^i = \left( N_{th}^i / N_{th}^{MB}\right) N_{exp}^{MB} $, essentially normalizing the events in the i-experiment to match the MiniBooNE excess. 

    \item To simultaneously satisfy both of the above conditions, we need to select the model's parameter space in a manner that the upscattering cross section is predominantly governed by the scalar-mediated contribution, and this cross section should decrease as the neutrino energy increases.
\end{enumerate}

Under the assumption that the heavy neutrino, as well as both the scalar and vector mediators, have short lifetimes, we can express the total number of events as follows:
\begin{equation}    
N^{MB}_{th} = \xi_{exp} \int_{{E_\nu}_{min}}^{{E_\nu}_{max}} d E_\nu \Phi(E_\nu) \int_{{E_R}_{min}}^{{E_R}_{max}} d E_R \frac{d \sigma (E_R, E_\nu)}{d E_R} \times Br\left( mediator \rightarrow e^+ e^- \right), \end{equation} 
where $\xi_{exp}$ accounts for the exposure time of the detector to the beam, the number of protons on target, and the effective area of the detector, and it depends on the specific experiment; $E_\nu$ is the energy of the incoming $\nu_\mu$ beam; $\Phi(E_\nu)$ is the corresponding flux; $E_R$ is the recoil energy of the target nuclei in the detector. To compare with the previous MiniBooNE solutions, we just need to calculate $\xi_{model} = N^{MB}_{th}/\xi_{exp}$.

The differential scattering cross-section with respect to the recoil energy $E_R$ for $\nu A \rightarrow \nu_s A$  mediated through the light scalar $h_1$ shown in upper panel of figure.~\ref{fig:MB} can be expressed as follows: 
\begin{equation}
    \frac{d \sigma}{d E_R} = \frac{y_\nu^2}{16 \pi E_\nu^2} \times \left[ Z f_p +(A-Z) f_n \right]^2 \times \frac{\left( {m^2_{\nu_{s_2}}} + 2 M_A E_R\right) \left(2 M_A + E_R \right)}{\left( m_{h_1}^2 + 2 M_A E_R \right)^2} \times F^2(E_R)
\end{equation} where $m_A$ is the mass of the target nucleus; $Z$ is the number of protons in the target nucleus; $A - Z$ represents the number of neutrons in the target nucleus, where $A$ is the total number of nucleons; $F({E_R})$ is the nuclear form factor~\cite{Helm:1956zz, Engel:1991wq}, which characterizes the spatial distribution of nucleons within the target nucleus and depends on the momentum transfer. The coupling factors $f_{p,n}$ are expressed as~\cite{Falk:1999mq} \begin{equation}  \frac{f_{p,n}}{m_N} = \sum_{q=u,d,s} f^{(p,n)}_{T_q} \frac{f_q}{m_q} + \frac{2}{27} \left( 1- \sum_{q=u,d,s} f^{(p,n)}_{T_q} \right) \sum_{q=c,b,t} \frac{f_q}{m_q}\end{equation} where we set $f_{(u,d)}= y_{(u,d)h_1}$ and $f_{c,s,t,b} =0$; the parameters $f_{T_q}^{(N)}$ are defined as $ m_N f_{T_q}^{(N)} = \braket{N|m_q \bar{q} q|N} $ for $N=p,n$~\cite{Shifman:1978zn}; our results used the values: $f_{T_u}^{(p)} = 0.020$, $f_{T_d}^{(p)} = 0.041$,$f_{T_u}^{(n)} = 0.0189$, and $f_{T_d}^{(n)} = 0.045$~\cite{Alarcon:2011zs, Alarcon:2012nr, Crivellin:2013ipa, Hoferichter:2015dsa, Junnarkar:2013ac}.

\begin{figure}[tbp]
\centering
\includegraphics[height=9cm,width=14cm]{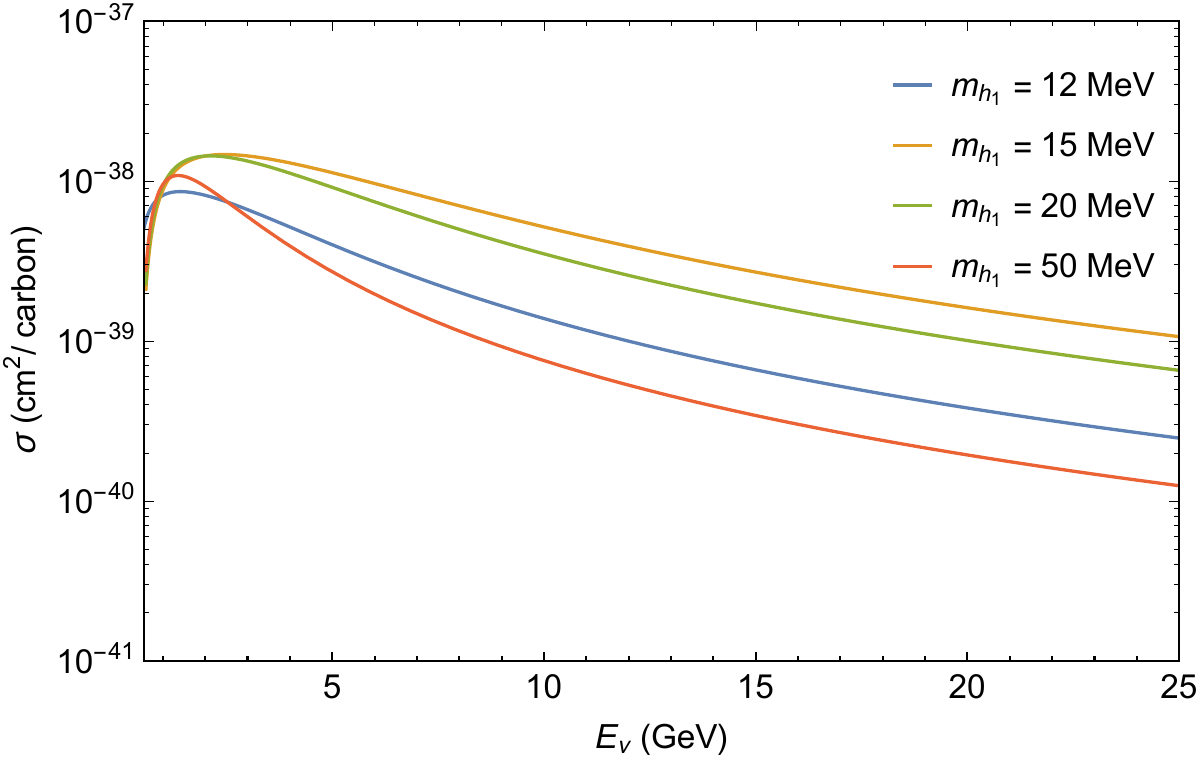}
\captionsetup{justification   = RaggedRight,
             labelfont = bf}
\caption{\label{fig:MBCross} This figure illustrates the cross sections' dependency on the incoming neutrino energy for various scalar masses in conjunction with a 17 MeV vector gauge boson. The  lines represent the combined cross section for both scalar and vector interactions for the benchmark points given in table.~\ref{table:MBbenchmark}.
 }
\end{figure}

The expression for the differential scattering cross-section mediated by the light vector $Z^\prime$ given by the two figures in the bottom panel of figure.~\ref{fig:MB} is given by,  
\begin{eqnarray}
   \frac{d \sigma}{d E_R} &=& \frac{g_\nu^2}{4 \pi E_\nu^2} \times \left[ (A+Z) C_{u,V} + (2A - Z) C_{d,V} \right]^2 \times F^2(E_R) \nonumber\\ & \times &\frac{ \left( 2 M_A E_\nu^2 + M_A E_R^2 - 2 M_A E_\nu E_R - {m^2_{\nu_{s_2}}} E_\nu + \frac{{m^2_{\nu_{s_2}}}}{2} E_R - \frac{{m^2_{\nu_{s_2}}}}{2} M_A- M_A^2 E_R \right) }{\left( m_{Z^\prime}^2 + 2 M_A E_R\right)^2}  \end{eqnarray} where $g_\nu$ represents the effective coupling between the active and sterile neutrinos with the vector gauge boson. This coupling is small due to the small mixing angle in the neutrino sector. It is  worth noting that the mixing angle is essentially a free parameter in our model since we are not making any other adjustments that would affect it.

\begin{table}[tbp]
\centering
\begin{tabular}{ lllll }
\hline
Parameters & BP1 &BP2 & BP3 & BP4 \\
&&&& \\ \hline
$m_{h_1}$& 12 MeV & 15 MeV & 20 MeV & 50 MeV \\ 
&&&& \\
$m_{\nu_{s_2}}$& 420 MeV & 420 MeV& 420 MeV & 420 MeV \\
&&&& \\

$y_\nu$& 0.0096 & 0.0182& 0.0190 & 0.0220 \\
&&&& \\
 $y_{uh_1}$& $5 \times 10^{-6}$ &  $5 \times 10^{-6}$ & $5 \times 10^{-6}$ & $5 \times 10^{-6}$ \\
 &&&& \\
 $y_{dh_1}$& $ 10^{-5}$ &  $ 10^{-5}$ & $ 10^{-5}$ & $ 10^{-5}$\\ 
 &&&& \\
$N_{MB}^{th}$& 636 & 647 &681 & 658\\ 
&&&& \\
\hline
\end{tabular}
\captionsetup{justification   = RaggedRight,
             labelfont = bf}
\caption{ \label{table:MBbenchmark} We present four benchmark scenarios with varying masses of the scalar boson that successfully account for the observed number of excess events at the MB detector. Importantly, these scenarios do not lead to an excessive production of events at the ND-T2K and MINERvA detectors. We have set the vector boson mass to $m_{Z^\prime} = 17$ MeV, while the gauge couplings are held constant at $g_X = 1.8 \times 10^{-3}$ and $g_X^\prime = 0.41 \times 10^{-4}$.}

\end{table}
To estimate the number of events predicted by our model, denoted as $N^{th}_{MB}$, we consider four benchmark points detailed in Table.~\ref{table:MBbenchmark}. Specifically, we explore four different scalar masses ranging from 10 MeV to 50 MeV, while keeping the masses of $\nu_{s_2}$ and $Z^\prime$ fixed at 420 MeV and 17 MeV, respectively. For the $Z^\prime$-mediated diagram, we choose the couplings to be $g_X = 1.8 \times 10^{-3}$ and $g_X^\prime = 4.1 \times 10^{-5}$. It is important to note that the neutrino coupling with $Z^\prime$ is further suppressed by  mixing angle in the neutrino sector. Consequently, the contribution to the cross-section from the $Z^\prime$-mediated diagram is orders of magnitude smaller compared to the scalar-mediated diagram. Thus, the total cross-section is predominantly determined by the scalar-mediated process. Concerning the scalar-mediated diagram, the quark couplings are constrained by the quark masses. However, the neutrino coupling $y_\nu$ is a free parameter and can be relatively large. Assuming typical neutrino and scalar energies of $E_{\nu_{s_2}}$ and $E_{h_1}$ around 1 GeV, we estimate that the length of the path they travel before decaying is less than or equal to $10^{-4}$ m, indicating prompt decays. In a related study~\cite{Brdar:2020tle}, it was demonstrated that a process dominated by scalar interactions does not lead to an overproduction of events at the T2K ND280 and MINER$\nu$A experiments, provided that $\nu_{S_2}$ decays promptly with a decay length of less than or equal to $10^{-4}$ meter.

\section{Summary and discussions} \label{sec:summary}

In light of the absence of evidence for new physics at the TeV scale in LHC experiments, attention is turning towards low-energy experiments that may provide insight into new physics phenomena at the MeV scale. Several intriguing findings have emerged recently from such experiments, including the Beryllium anomaly at the ATOMKI experiment, the anomalous magnetic moment of the muon, and the excess low-energy events observed at MiniBooNE. To address these anomalies simultaneously, we propose a straightforward model extension of the gauge and scalar sectors  within the SM. This extension introduces new scalar fields $H_2$ and $\phi_H$ that carry charges under a new gauge group, denoted as $U(1)_H$. The model is $U(1)_H$ extensions of the Type-I 2HDM plus a singlet scalar $\phi_H$.  The mixing within the scalar sector yields a new light scalar field, while mixing within the gauge sector results in a light vector boson. Additionally, our model incorporates $U(1)_H$-charged sterile neutrinos at the MeV scale.

The resolution of the Beryllium anomaly primarily relies on the presence of the vector boson mass around 17 MeV, whereas the excess events observed at MiniBooNE can be predominantly attributed to the scalar particle. Notably, both the scalar and vector particles contribute to the muon $(g-2)$, albeit with opposite signs. Careful fine-tuning between these contributions allows us to arrive at the correct value for the muon's magnetic moment. Hence, by utilizing the capabilities of the two light mediators within our model, we simultaneously provide explanations for these recent significant anomalies.

These results will soon undergo verification through upcoming experiments. Notably, MEG II at PSI is set to replicate the Beryllium experiment conducted at ATOMKI, E34 at J-PARC aims to validate our model's predictions regarding the muon's magnetic moment, and MicroBooNE is actively scrutinizing the MiniBooNE results for further insights

\acknowledgments

  This work is  supported in part by KIAS Individual Grants under Grant No. PG021403 (PK), and by National Research Foundation 
  of Korea (NRF) Grant No. NRF-2019R1A2C3005009 (SG PK). 
  We acknowledge that we have used the TikZ-Feynman~\cite{Ellis:2016jkw} package to generate the Feynman diagram of Figs.~\ref{fig:g2fig} and \ref{fig:MB}.

\bibliographystyle{JHEP.bst}
\bibliography{ref.bib}

\end{document}